\documentclass[]{JHEP3}
\pdfoutput=1

\usepackage{graphicx}
\usepackage{amsmath,amssymb,amscd,amsfonts,bm}

\usepackage{color}

\def\<>#1{\big\langle{#1}\big\rangle}
\def\[]#1{\big[{#1}\big]}

\title{Combining subjet algorithms to enhance $ZH$ detection at the LHC}

\author{Davison E. Soper\\
Institute of Theoretical Science\\
University of Oregon\\
Eugene, OR  97403-5203, USA\\
E-mail: \email{soper@uoregon.edu}
}
\author{Michael Spannowsky \\
Institute of Theoretical Science\\
University of Oregon\\
Eugene, OR  97403-5203, USA\\
E-mail: \email{mspannow@uoregon.edu}
}

\abstract{The signal for a highly boosted heavy resonance competing against a background of light parton jets at the LHC can be enhanced by analyzing subjets in the ``fat'' jet that possibly contains the heavy resonance. Three methods for doing this are known as filtering, pruning, and trimming. We study the possibility of combining these methods using a relative likelihood approach. We find that, because the methods are not the same, one achieves an enhanced statistical power by combining them. We illustrate the possibilities first with a simple problem of combining trimming and pruning to enhance the signal for finding a boosted top quark. We then study the more difficult problem of disentangling from the background the signal for the production of a Higgs boson in association with a $Z$-boson. For this problem, we combine filtering, trimming, and pruning.
}

\preprint{ }

\begin{document}


\section{Introduction}

A central problem for data analysis at the Large Hadron Collider (LHC) is to find the signal for the production of a new heavy particle against a background of jets produced by standard model processes that do not involve the heavy particle. An important example is the production of a Higgs boson in association with a $Z$-boson, where the $Z$-boson decays into $e^+ e^-$ or $\mu^+ \mu^-$ and the Higgs boson decays into $b \bar b$. One can demand that the $Z$-boson has large transverse momentum, say $P_T > 200\ {\rm GeV}$. Since it recoils against the $Z$-boson, the Higgs boson then has a large transverse momentum and is easier to find than if it had low transverse momentum \cite{Butterworth:2008iy}. Nevertheless, there is a large background to this process from standard model processes that do not involve the Higgs boson, so some ingenuity is required to separate the signal from the background.

Three methods have been proposed for this sort of problem: filtering \cite{Butterworth:2008iy}, trimming \cite{Krohn:2009th}, and pruning \cite{Ellis:2009su,Ellis:2009me}. These methods are not the same and, because they are not the same, there is more information available in combinations of two or three of them than is available in any one method. We will illustrate this in this paper. We will find that, by combining methods, we can enhance the significance of the signal. 

There are other cases of searches for as yet undiscovered physics in which the easiest sort of data analysis would involve simply looking for a bump in the mass distribution of jets found with the $k_T$ jet algorithm or a similar standard algorithm. In such cases, filtering, trimming, or pruning the jet can help, but may still produce a barely sufficient statistical significance for finding the signal with the integrated luminosity that is available. In such cases, combining methods as described in this paper can further enhance the signal.

We combine methods using a likelihood analysis. We will also see that methods can be combined using a cut-based analysis, but the likelihood method is more powerful.

The analysis of this paper applies in general to processes in which the signal for a highly boosted ({\it i.e.} high transverse momentum) heavy resonance competes against a background of light parton ($g,u,\bar u, d, \bar d, s, \bar s, c,\bar c, b, \bar b$) jets. We begin in section \ref{sec:top} with a simple example in which we combine trimming and pruning to enhance the signal for a highly boosted top quark that decays into hadrons. This example gives us a chance to outline briefly what trimming and pruning are and to then illustrate how they can be combined. We then turn, in section \ref{Sec:ZH}, to the more challenging $ZH$ production process. Here we use filtering, trimming, and pruning. Some conclusions follow in section \ref{sec:conclusion}. In appendix \ref{sec:Likely} we review some basics of the likelihood analysis that we use.

\section{Top quark identification} 
\label{sec:top}

This paper concerns detecting Higgs boson production in association with a $Z$-boson. Our analysis of $ZH$ production involves three methods of subjet analysis, known as trimming, pruning, and filtering, the last with tagging of subjets containing a $b$-quark. In addition, we use the likelihood ratio both as a measure of statistical significance and as a tool for combining different methods for detecting the same signal. In order not to introduce too many ideas at once in a somewhat complex analysis, we choose to introduce  in this section some of the needed concepts: trimming, pruning, and our use of the likelihood ratio.  We introduce these tools with the aid of a simple but quite artificial analysis involving the identification of a top-quark jet. We note that the reconstruction of boosted top jets has been considered in many different subjet analysis before \cite{Ellis:2009su,Ellis:2009me,Brooijmans:2008zz, Kaplan:2008ie, Thaler:2008ju, Almeida:2008tp,Krohn:2009wm,Plehn:2009rk}. We emphasize that the ``signal'' that we study is not realistic, nor do we apply sensible cuts to define the signal and background event samples. Rather, we analyze ``signal'' and ``background'' event samples only to introduce some of the conceptual ingredients that we need. Once we have seen these ingredients, we can turn to $ZH$ production in section \ref{Sec:ZH}.

For this study, we use AlpGen \cite{Mangano:2002ea} to generate ``signal'' and background events and we shower the events using Pythia 6.3 \cite{Sjostrand:2006za} and recombine the jets using FastJet \cite{Cacciari:2005hq}. We have in mind a scenario where a resonance in the s-channel ({\it e.g.} from a strongly coupled sector \cite{Evans:2009ga}) splits into $t\bar{t}$. To mimic the signal of this scenario, we generate events with $t\bar t$ pairs in which the $t$ and the $\bar t$ have large transverse momenta. We use the standard model process $gg \to t\bar t$ to generate the top quarks, but with half the standard model differential cross section and with a cut $P_T > 350\ {\rm GeV}$ for the hardest of the $t$ or $\bar t$. We select events in which one of the top quarks decays to $b \ell \nu$, $\ell = e$ or $\mu$, with $P_{T,\ell} > 15~\mathrm{ GeV }$. The other top quark decays hadronically, to $b q \bar q$. The total signal cross-section that we generate with these cuts is 1.5 pb. We generate the background events using standard model production of a $W$-boson (with $W \to \ell \nu$) recoiling against light parton jets. Events are selected for the analysis if the largest $P_T$ jet (with the inclusive Cambridge-Aachen algorithm \cite{CAjet} with $R = 1.5$) has $P_T > 100\ {\rm GeV}$. Here and throughout this paper we take $\sqrt s = 14\ {\rm TeV}$.

Having generated events, we now set the problem: to find the top quark that decays hadronically against the background of light parton jets that recoil against a $W$-boson.\footnote{If we really wanted to do a good job of finding the signal that we have chosen, we would impose a stiff $P_T$ cut on all jets chosen for analysis. For our pedagogical purposes, we choose not to do this.} To start, we find the largest $P_T$ jet in the event.\footnote{Jets are considered only if the absolute value of their rapidity $y$ is less than 5. This is a very non-restrictive cut. However the highest $P_T$ jet is quite likely to have $|y|$ much less than 5.} This jet should contain the decay products of the top quark if there is a top quark. To make sure that the decay products are well contained, we should use a jet finding algorithm that uses a fairly wide angular range. There is some choice here. We use either the inclusive Cambridge-Aachen algorithm \cite{CAjet} with a cone size $R = 1.5$ or the anti-$k_T$ algorithm \cite{antiKT} with $R = 1.5$. Because of the large angular size used in the jet finding algorithm, we call this the fat jet.

The simplest way to proceed from here would be to measure the invariant mass $M_{\rm Jet}$ of the fat jet, expecting to find $M_{\rm Jet} \approx M_{\rm top} = 174\ {\rm GeV}$. However, jets from the background sample with this angular size can have large masses. Thus we expect that the distribution of $M_{\rm Jet}$ for background events will be substantial around the region of interest, $M_{\rm Jet} \approx M_{\rm top}$. Furthermore, we cannot expect the signal events to yield a narrow peak near $M_{\rm Jet} = M_{\rm top}$ because the fat jet will inevitably contain hadrons from partons that originate in initial state radiation and from secondary interactions in the underlying event. These extra hadrons add to $M_{\rm Jet}$ and thus smear the signal distribution.\footnote{Indeed, there is not even a clear distinction between partons radiated from the initial state and from the top quark and its daughters because the quantum amplitudes that represent these two sources can interfere.} For these reasons, we need to break the fat jet into subjets and analyze the structure of the subjets.

Consider first the trimming method \cite{Krohn:2009th}. Here, following Ref.~\cite{Krohn:2009th}, we define the fat jet using the anti-$k_T$ algorithm with $R = 1.5$. The fat jet is made of constituents that we can take to be individual hadrons or else very narrow jets made from calorimeter towers. Let us call them the starting protojets. We now apply a sequential clustering algorithm to the protojets, grouping them into successively fatter protojets. There is a choice of algorithm to use. We use the $k_T$ algorithm \cite{KTjet} with protojet recombination defined by adding the four-momenta of the protojets. This algorithm has an effective cone size $R$ and here we choose a quite small cone, $R = 0.2$. After the $k_T$ algorithm has combined the starting protojets up to a $k_T$ limit defined by this $R$, we have a list of jets, each consisting of some subset of the original starting protojets. There may be, say, ten final jets. We are ready to trim our list of jets, keeping relatively hard jets and throwing away relatively soft jets. We keep jet $j$ if
\begin{equation}
\label{eq:trimmingkeep}
P_{T,j} > f \times \Lambda
\
\;\;,
\end{equation}
were the hard scale $\Lambda$ is the $P_T$ of the fat jet. The fraction $f$ is an adjustable parameter that we take to be $f = 0.03$. The starting protojets $i$ contained in the jets $j$ for which the inequality (\ref{eq:trimmingkeep}) holds constitute the trimmed jet. Now we measure the invariant mass of the trimmed jet,
\begin{equation}
M_{\rm Jet}^2 = \left(\sum_{i} p_i \right)^2
\;\;.
\end{equation}

For background events, trimming reduces $M_{\rm Jet}$ for each event and thus reduces the high $M_{\rm Jet}$ part of the jet-mass distribution. For signal events, trimming removes extraneous parts of the jets, giving a sharper peak near $M_{\rm Jet} = M_{\rm top}$. The result is illustrated in figure \ref{fig:trimming}. The $\bar t t$ signal is clearly visible. We will investigate the statistical significance of the signal shortly.

\FIGURE{
\includegraphics[width=9.0cm]{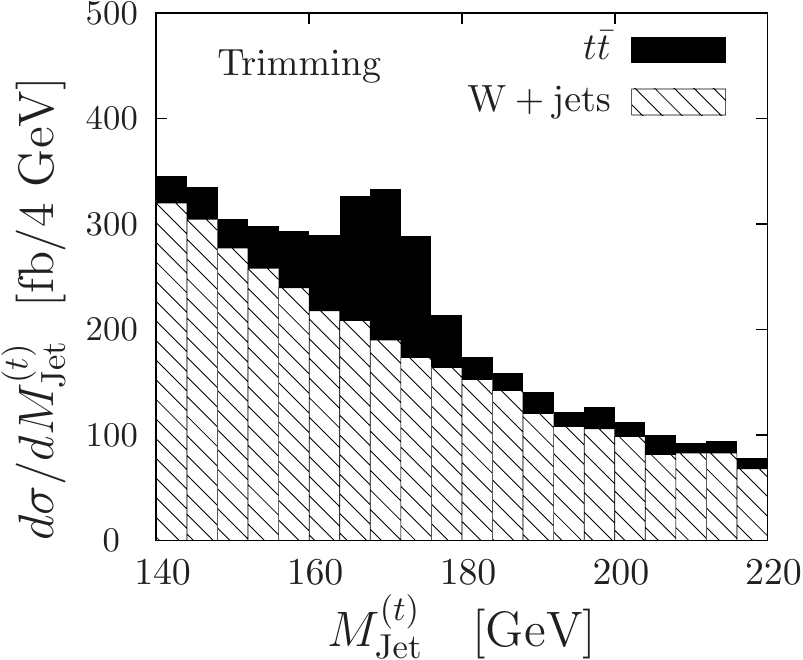}
\caption{Mass distribution of trimmed jets for the $t\bar t$ signal and the $W + {\rm jets}$ background. The top mass is taken to be $174~\mathrm{GeV}$.  }
\label{fig:trimming}
}

Now consider the pruning method \cite{Ellis:2009su,Ellis:2009me}. Here, following Refs.~\cite{Ellis:2009su,Ellis:2009me}, we define the fat jet using the inclusive Cambridge-Aachen algorithm with $R = 1.5$. We again take the fat jet to be composed of very narrow starting protojets and apply a sequential clustering algorithm to the protojets. This time, we choose a modified version of the Cambridge-Aachen algorithm \cite{CAjet}.\footnote{Other successive combination jet algorithms are allowed for the pruning method. With the use of the Cambridge-Aachen algorithm, the description of how the method works is somewhat simplified.} In this algorithm, in each step we look for the pair of protojets $\{i,j\}$ for which 
\begin{equation}
R_{i,j}^2 = (y_i - y_j)^2 + (\phi_i - \phi_j)^2
\end{equation}
is the smallest. (Here $y_i$ is the rapidity of protojet $i$ and $\phi_i$ is its azimuthal angle.) This pair of protojets is combined by adding their four-momenta, creating a new protojet. The normal Cambridge-Aachen algorithm continues until no pair $\{i,j\}$ of protojets has $R_{i,j} < D_{\rm cut}$, where $D_{\rm cut}$ is a parameter that represents an effective cone size for this algorithm. We take $D_{\rm cut} = M({\rm fat\ jet})/P_T({\rm fat\ jet})$ and let the algorithm run until it stops. At this stage, each pair $\{i,j\}$ of protojets has $R_{i,j} > D_{\rm cut}$. Now we let protojet combination continue, but with an additional restriction: for each pair $\{i,j\}$ of protojets that are ready to be combined, we look at the momentum fraction
\begin{equation}
\label{eq:zdef}
z = \frac{\min(p_{T,i},p_{T,j})}{|\vec p_{T,i} + \vec p_{T,j}|}
\;\;.
\end{equation}
If $z$ is small, there is a danger that we are including a protojet that is extraneous to the signal. Therefore, if 
\begin{equation}
\label{eq:zcutdef}
z < z_{\rm cut}
\;\;,
\end{equation}
we do {\rm not} combine protojets $i$ and $j$ and instead simply drop whichever of the two protojets had the smaller transverse momentum. For this $t\bar t$ analysis, we take $z_{\rm cut} = 0.1$. Then we continue with the algorithm until all protojets have either been combined or else eliminated. The starting protojets contained in the final jet when the algorithm stops constitute the pruned jet. Now we measure the invariant mass of the pruned jet,
\begin{equation}
M_{\rm Jet}^2 = \left(\sum_{i} p_i \right)^2
\;\;.
\end{equation}
The motivation for pruning is essentially the same as for trimming. For background events, pruning reduces $M_{\rm Jet}$ for each event and thus reduces the high $M_{\rm Jet}$ part of the jet-mass distribution. For signal events, pruning removes extraneous parts of the jets, giving a sharper peak near $M_{\rm Jet} = M_{\rm top}$. The result is illustrated in figure \ref{fig:pruning}. The $\bar t t$ signal is visible, perhaps less so than with trimming. 

\FIGURE{
\includegraphics[width=9.0cm]{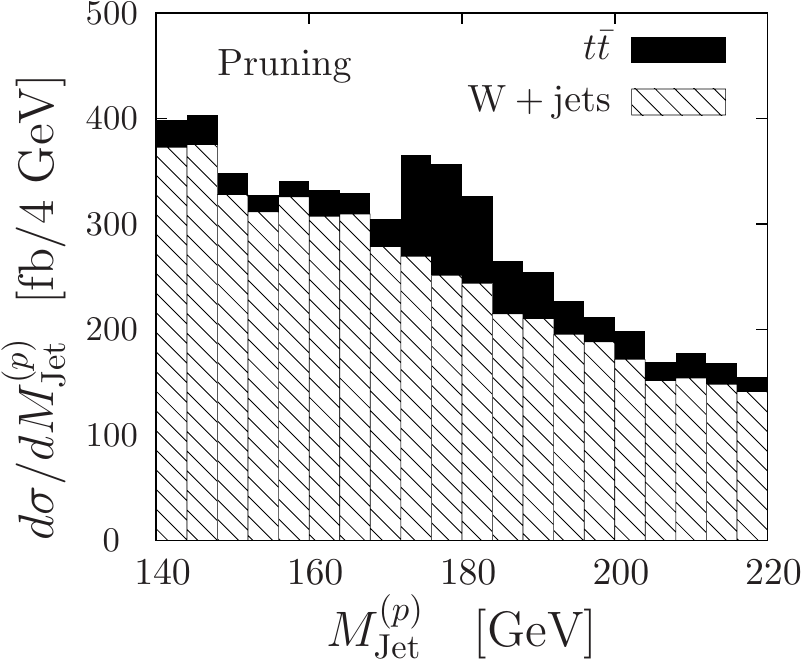}
\caption{Mass distribution of pruned jets for the $t\bar t$ signal and the $W + {\rm jets}$ background. The top mass is taken to be $174~\mathrm{GeV}$.}
\label{fig:pruning}
}

In figures \ref{fig:trimming} and \ref{fig:pruning}, the $t \bar t$ signal is obvious. However, these figures represent theoretical cross sections. Data will look different. The simplest way to look for the signal in data is to define a mass window and count the events in the window. Let us take our mass window to be $160\ {\rm GeV} < M_{\rm Jet} < 200\ {\rm GeV}$. Let $n$ be the number of events in this window after a certain amount of integrated luminosity has been accumulated. We will take the integrated luminosity to be $\int\!dL = 30\ {\rm pb}^{-1}$. Let $b$ be the expectation value of the number of background events with this luminosity and let $s$ be the expected number of signal events. That is, $b$ and $s$ are the theoretical cross sections times $\int\!dL$. 

To assess what we learn from the measurement in the face of counting statistics, we consider that there are two competing interpretations of the data: that it all arises from the $W + {\rm jets}$ background (theory B) or that it arises from this background plus the $t\bar t$ signal (theory SB). The probability that $n$ events are measured if there is only background is $b^n e^{-b}/n!$. The probability that $n$ events are measured if there is a signal plus the background is $(b+s)^n e^{-(s+b)}/n!$. The ratio of these, known as the likelihood ratio, is $\exp({\cal L})$ where
\begin{equation}
{\cal L}(n) = n \log\left(1 + \frac{s}{b}\right) - s
\;\;.
\end{equation}
If ${\cal L}$ is substantially greater than 1, the result strongly favors the interpretation that the $t\bar t$ signal is present. For instance ${\cal L} = 4$  favors the presence of the signal by a ratio $\exp({\cal L}) \approx 55$. We review some properties of the likelihood ratio in appendix \ref{sec:Likely}.

The expectation value of ${\cal L}(n)$ if the SB theory is right is
\begin{equation}
\label{eq:expectedA}
\langle {\cal L}(n)\rangle_{\rm SB} = (s+b) \log\left(1 + \frac{s}{b}\right) - s
\;\;.
\end{equation}
Thus, we can expect to reliably see the $t\bar t$ signal if $\langle {\cal L}(n)\rangle_{\rm SB}$ is substantially greater than 1. As a minimum requirement, we may ask for $\langle {\cal L}(n)\rangle_{\rm SB} > 4$. The results are shown in table \ref{tab:trimorprune}. We see that trimming does better than pruning, but neither method provides enough statistical power to achieve $\langle {\cal L}(n)\rangle_{\rm SB} > 4$ with an integrated luminosity of just $30\ {\rm pb}^{-1}$. (Of course, the statistical insufficiency goes away with more luminosity, but in this simple example we imagine that $30\ {\rm pb}^{-1}$ is all the luminosity that we have.)

\TABLE{
\begin{tabular}{c|c|c}
  & Trimming & Pruning \\ 
	\hline
Signal cross section [fb]        &  590  &  503           \\	
Background cross section [fb]   &  1571  &  2480          \\	
$s/b$                           &   0.38   &   0.20          \\
$s/\sqrt{b}$ \ \ ($\int\!dL = 30\ {\rm pb}^{-1}$) &  2.6 & 1.7 \\
$\langle{\cal L}(n)\rangle_{\rm SB}$  \ \ ($\int\!dL = 30\ {\rm pb}^{-1}$) & 3.0 & 1.4
\end{tabular}
\caption{Statistical significance of trimming and pruning results for an integrated luminosity  of $30\ {\rm pb}^{-1}$. Here we simply count the expected number of signal events, $s$, and background events, $b$, in a top quark mass window $160\ {\rm GeV} < M_{\rm Jet} < 200\ {\rm GeV}$. The logarithm of the likelihood ratio based on these expected counts is $\langle{\cal L}(n)\rangle_{\rm SB}$, eq.~(\ref{eq:expectedA}).}
\label{tab:trimorprune}
}

It is rather artificial to base the SB {\em vs.} B assessment on simply the counts in a single jet mass window. The experiment will give counts $n_J$ in each bin $J$ shown in figures \ref{fig:trimming} and \ref{fig:pruning}. We can base our assessment on the log likelihood ratio using all of the information. Then the likelihood ratio is the product of the likelihood ratios for all of the bins used. Its logarithm is
\begin{equation}
\label{eq:loglikelihooddef}
{\cal L}(\{n\}) = \sum_J 
\left[n_J \log\left(1 + \frac{s_J}{b_J}\right) - s_J\right]
\;\;.
\end{equation}
Here $n_J$ is the number of events in bin $J$ and $s_J$ and $b_J$ are the corresponding signal and background cross sections times the integrated luminosity.

The expectation value of ${\cal L}(\{n\})$ if the SB theory is right is
\begin{equation}
\label{eq:expectedWfull}
\langle {\cal L}(\{n\})\rangle_{\rm SB} = 
\sum_J 
\left[(s_J + b_J) \log\left(1 + \frac{s_J}{b_J}\right) - s_J\right]
\;\;.
\end{equation}
Using the full bin-by-bin information, we find
\begin{equation}
\begin{split}
\label{eq:loglikelihoodtrimorprune}
\langle {\cal L}(\{n\})\rangle_{\rm SB} ={}& 4.4\,,
\hskip 1 cm {\rm trimming}\;,
\\
\langle {\cal L}(\{n\})\rangle_{\rm SB} ={}& 2.4\,,
\hskip 1 cm {\rm pruning}\;\;.
\end{split}
\end{equation}
We see that using the full available information improves the discriminating power of the experiment. In fact, now the log likelihood ratio with trimming is above our nominal threshold of 4.0.

Although trimming and pruning are rather similar in spirit, they are different. This is illustrated in figure \ref{fig:trimvsprune}. In the left-hand plot, we evaluate the trimmed jet mass, $M_{\rm Jet}^{(t)}$, and the pruned jet mass, $M_{\rm Jet}^{(p)}$, for each simulated $t\bar t$ signal event. We accumulate events in bins of $\{M_{\rm Jet}^{(t)},M_{\rm Jet}^{(p)}\}$ and plot the resulting density of events. We see from the plot that the bins with the most events do not lie along the diagonal, $M_{\rm Jet}^{(t)} = M_{\rm Jet}^{(p)}$, in this plot. In fact, the bins with the most events have $M_{\rm Jet}^{(t)} < M_{\rm Jet}^{(p)}$. In the right hand plot, we do the same thing for the background events. Again, the most populated bins do not lie along the diagonal (or along any one-dimensional curve). 

\FIGURE{
\includegraphics[width=7.0cm]{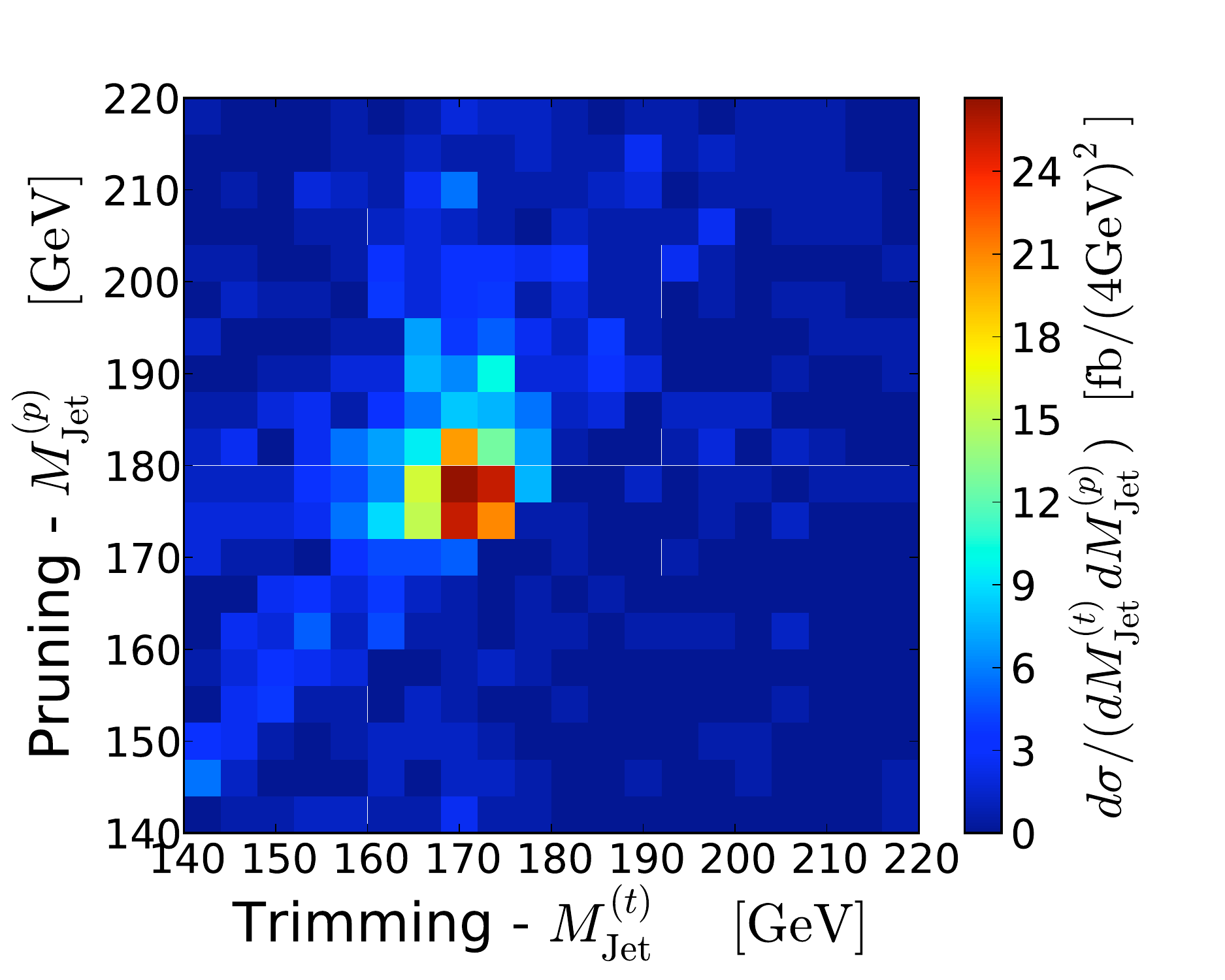}
\includegraphics[width=7.0cm]{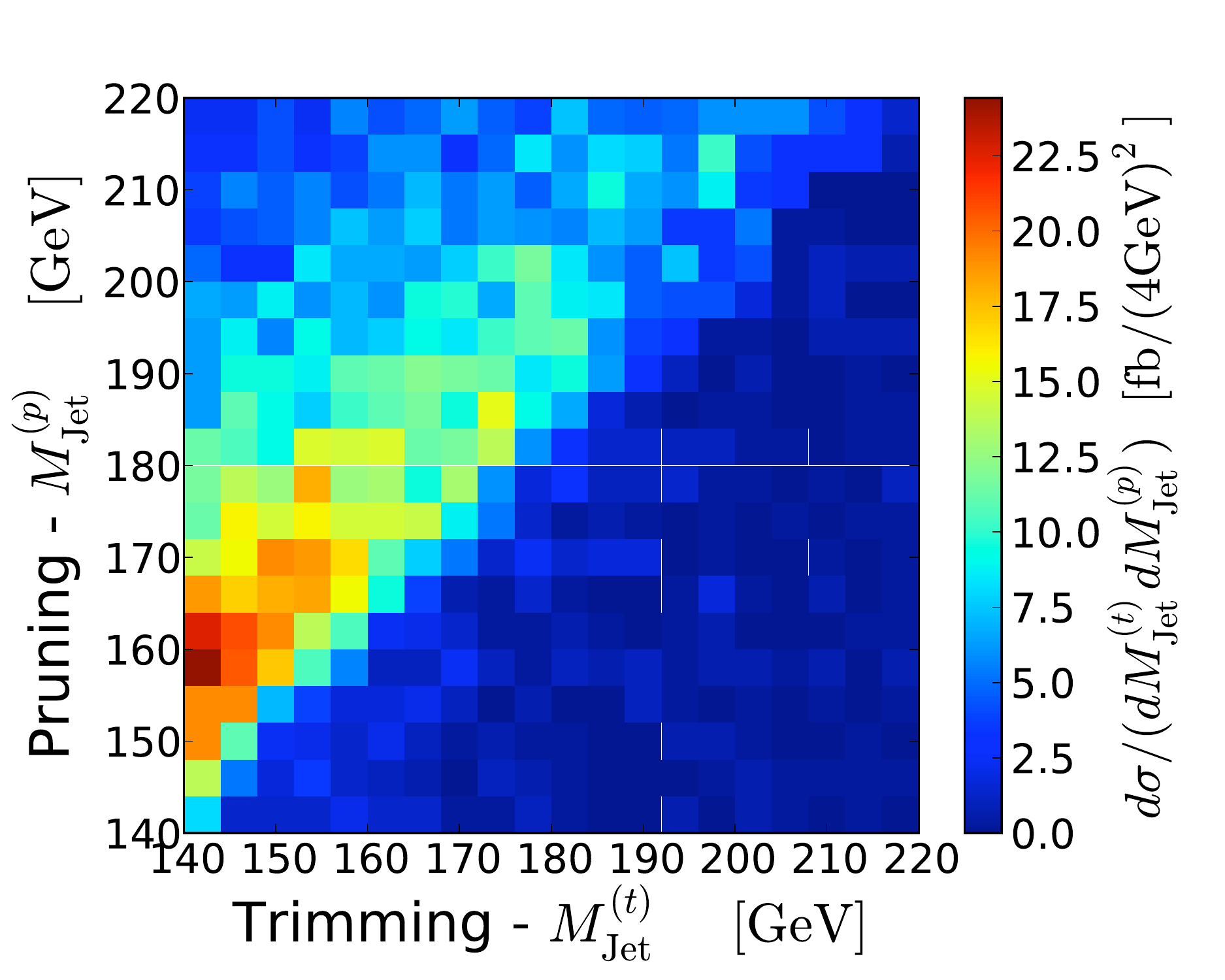}  
\caption{Trimmed jet mass and pruned jet mass in the $t \bar{t}$ production process and the background process $W + {\rm jets}$. The left plot shows the signal for $M_{\rm top}=174~{\rm GeV}$; the right plot shows the background. Generally, trimmed jet masses are smaller than pruned jet masses, but there is no fixed relation between them.}
\label{fig:trimvsprune}
}

Because trimmed jet masses and pruned jet masses contain different information, it may be possible to obtain stronger results by using both of them. The simplest way to do this is to measure the number of events for which {\em both} masses fall into the top quark mass window $160\ {\rm GeV} < M_{\rm Jet} < 200\ {\rm GeV}$. In this case, we obtain the results summarized in table \ref{tab:trimandprune}. We see that there is some improvement in the statistical significance compared to the results in table \ref{tab:trimorprune}. Additionally, $s/b$ is larger when trimming and pruning are combined. This is important if the normalization of $b$ is not precisely known.

Evidently, we could also try to improve the statistical significance by adjusting the mass windows used for the trimmed and pruned jet masses. We do not, however, pursue this avenue.

\TABLE{
\begin{tabular}{c|c|c}
  &  Trimming  +  pruning  \\ 
	\hline
Signal cross section [fb]       &  360           \\	
Background cross section [fb]   &  508           \\	
$s/b$                           &  0.71          \\
$s/\sqrt{b}$ \ \ ($\int\!dL = 30\ {\rm pb}^{-1}$) &  2.8 \\
$\langle{\cal L}(n)\rangle_{\rm SB}$  \ \ ($\int\!dL = 30\ {\rm pb}^{-1}$) & 3.1
\end{tabular}
\label{tab:trimandprune}
\caption{Statistical significance of combined trimming and pruning results for an integrated luminosity  of $30\ {\rm pb}^{-1}$. Here we simply count events in which both trimmed jet mass and the pruned jet mass fall into the top quark mass window $160\ {\rm GeV} < M_{\rm Jet} < 200\ {\rm GeV}$. The notation is the same as in table \ref{tab:trimorprune}.
 }
}

Instead of combining the trimming and pruning information based on the event count in a single window, we use the log-likelihood ${\cal L}$, eq.~(\ref{eq:loglikelihooddef}), based on all of the bins in figure \ref{fig:trimvsprune} that contain a background cross section of at least $0.5\ {\rm fb}$.\footnote{The results are not sensitive to this cut, which we impose so that we can have a reliable calculation of $s_J/b_J$.} We find
\begin{equation}
\begin{split}
\langle {\cal L}(\{n\})\rangle_{\rm SB} ={}& 6.2\,,
\hskip 1 cm {\rm trimming} + {\rm pruning}
\;\;.
\end{split}
\end{equation}
This is a significant improvement on the log likelihood ratio that we obtained with either trimming or pruning alone, eq.~(\ref{eq:loglikelihoodtrimorprune}).

We can extend the analysis so as to display more information. The number of signal events in each bin is a function $s_J(m)$ of the top quark mass $m$ that we use to calculate the $t\bar t$ signal cross section. Until now, we have taken $m$ to be $M_{\rm top} = 174\ {\rm GeV}$. However, we can let $m$ vary. We consider the choices $m = (145,155,165,174,185,195,205,215)\, {\rm GeV}$. For each choice, we construct ${\cal L}(\{n\},m)$ according to eq.~(\ref{eq:loglikelihooddef}). Then, if we were to use data for the number of events $n_J$ in each bin, we would test not only whether the SB theory is favored over just the B theory, but also which values of $m$ are favored or disfavored by the data. To display what can be expected on average, we show in figure \ref{fig:trimandprune} the expectation value of ${\cal L}(\{n\},m)$ in the SB theory with the {\em true} top quark mass, $M_{\rm top} = 174 \ {\rm GeV}$. That is, 
\begin{equation}
\label{eq:expectedWfullwithmass}
\langle {\cal L}(\{n\},m)\rangle_{\rm SB} = 
\sum_J 
\left[(s_J(M_{\rm top}) + b_J) 
\log\left(1 + \frac{s_J(m)}{b_J}\right) - s_J(m)\right]
\;\;.
\end{equation}
The results are plotted in figure \ref{fig:trimandprune} as a function of $m$. We show the results for trimming alone, pruning alone, and for trimming and pruning combined. We see that the SB theory with $m = M_{\rm top}$ is highly favored, with a stronger result obtained if we combine trimming and pruning. We also see that the result using trimming and pruning combined is quite sensitive to the value of $m$: $m = M_{\rm top}$ is favored, while $m = 165\ {\rm GeV}$ and $m = 185\ {\rm GeV}$ are not favored. For these wrong values of $m$, $\langle {\cal L}(\{n\},m)\rangle_{\rm SB}$ is close to 0. For $m = 155\ {\rm GeV}$ and $m = 145\ {\rm GeV}$, the signal + background theory with the wrong $m$ is even weakly disfavored compared to the background only theory.

\FIGURE{
\includegraphics[width=10.0cm]{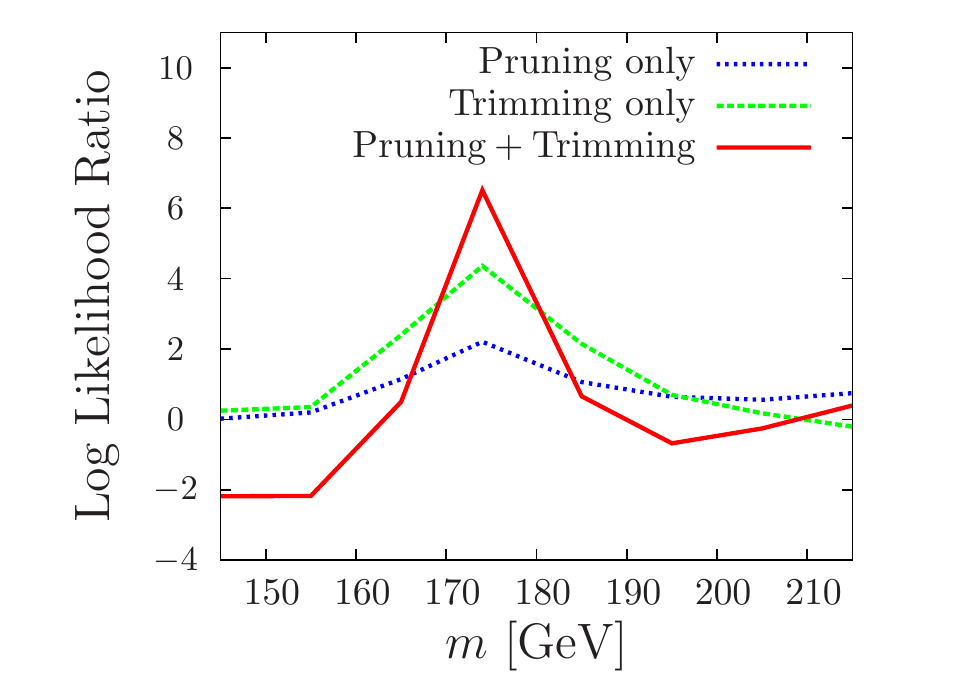}
\caption{The log likelihood ratio in $t \bar{t}$ production as a function of the trial top quark mass $m$, assuming an integrated luminosity of $30~\mathrm{pb}^{-1}$. We construct ${\cal L}(\{n\},m)$ for eight different values of $m$. Then we take the expectation value of these quantities in the signal + background theory with the true top mass, $M_{\rm top}$. The results are shown for trimming alone, pruning alone, and for trimming and pruning combined.}
\label{fig:trimandprune}
}
 
One should not think that figure \ref{fig:trimandprune} is what data will look like. We plot the expectation value of ${\cal L}(\{n\},m)$, but the values of the counts $n_J$ are subject to fluctuations. From appendix \ref{sec:Likely}, the variance of ${\cal L}(\{n\},m)$ is

\begin{equation}
\begin{split}
\big\langle ({\cal L} - \langle{\cal L}\rangle_{\rm SB})^2\big\rangle_{\rm SB} 
={}& 
\sum_{J} \big(b_j + s_J(M_{\rm top})\big) 
\left[\log\!\left(1 + \frac{s_J(m)}{b_J}\right)\right]^2
\;\;.
\end{split}
\end{equation}
Using the log likelihood results for trimming and pruning combined from figure \ref{fig:trimandprune}, we plot ${\cal L} \pm [\big\langle ({\cal L} - \langle{\cal L}\rangle_{\rm SB})^2\big\rangle_{\rm SB}]^{1/2}$ as an error band in figure \ref{fig:fluctuations}. Then we display five sample curves for ${\cal L}(\{n\})$ in which the counts $n_J$ in the bins $J$ are drawn from Poisson distributions with mean $s_J(M_{\rm top}) + b_J$. We see that the SB theory with the right mass is generally favored, but that it can be more or less favored depending on whether the counts in the most important bins fluctuate up or down.

\FIGURE{
\includegraphics[width=10.0cm]{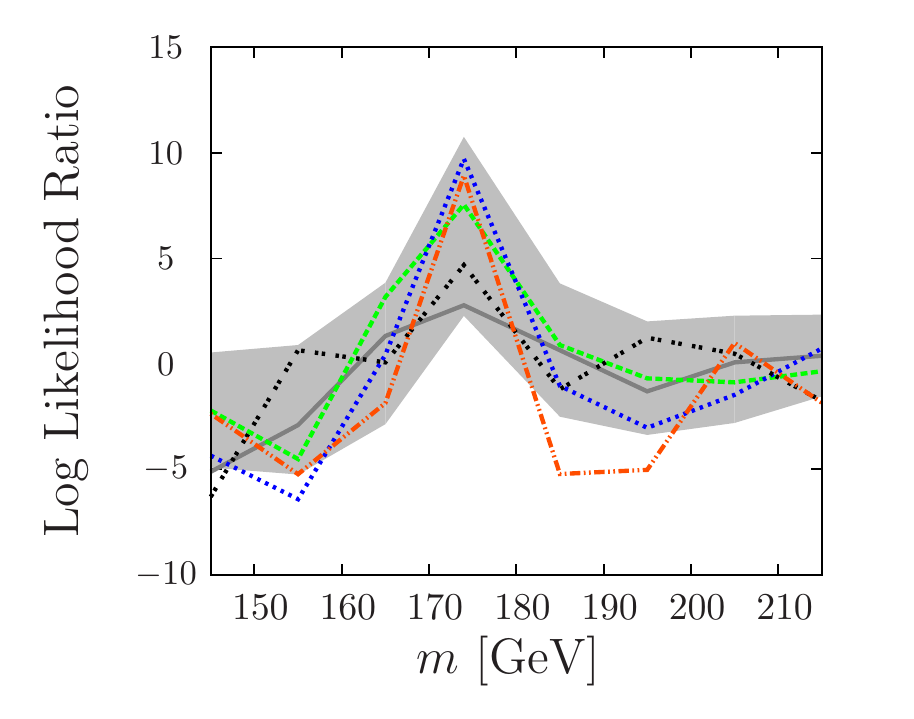}
\caption{Sample results for the log likelihood ratio in the $t \bar{t}$ production process as a function of the trial top quark mass $m$, using trimming and pruning combined and assuming an integrated luminosity of $30~\mathrm{pb}^{-1}$. We construct ${\cal L}(\{n\},m)$ for eight different values of $m$. Then we compute ${\cal L}(\{n\},m)$ for five random sets of the counts $n_J$ drawn from Poisson distributions with mean $s_J(M_{\rm top}) + b_J$. We also show an error band based on the mean value of ${\cal L}(\{n\},m)$ and its variance. Most points are within the error band, but note that 2 $\sigma$ or larger deviations either upward or downward will sometimes occur.}
\label{fig:fluctuations}
}

\section{$ZH$ production} \label{Sec:ZH}

We now turn to the production of a Higgs boson in association with a $Z$-boson, where the $Z$-boson decays into $e^+ e^-$ or $\mu^+ \mu^-$ and the Higgs boson decays into $b \bar b$. This process was considered in ref.~\cite{Butterworth:2008iy} and found to contribute to the overall signal significance for a Higgs boson search when $M_H \leq 130~\rm{GeV}$. The idea is to demand that the $Z$-boson have large transverse momentum. Then the recoiling Higgs boson has large transverse momentum and is easier to find against the backgrounds even though the cross section for this process is small.  The backgrounds that we consider are $ZZ$ production and, most importantly, $Z+\rm{jets}$ production. In the part of our analysis that uses the methods of ref.~\cite{Butterworth:2008iy}, we find good agreement with the results of ref.~\cite{Butterworth:2008iy}. Our purpose is to extend the analysis of ref.~\cite{Butterworth:2008iy} by investigating the improvement in background rejection obtained by using more than one algorithm for the analysis of subjets. 

The $HW$ production process is also important for a Higgs search, for the same reason as for $ZH$. However, this signal has an additional important background, $t\bar t$ production. In this paper, we restrict the analysis to the simpler $ZH$ case.

We generate the $ZH$, $ZZ$ and $Z+\rm{jets}$ samples using Pythia 8. We include an event in our sample if it has an electron or muon pair with 
\begin{align}
80~\mathrm{GeV} < m_{ll} & < 100~\mathrm{GeV}\;\;, \\
p_{T,ll} & > 200~\mathrm{GeV}\;\;.
\end{align}
The leptons are required to have rapidity $|\eta|<2.5$. We further require that there be no additional leptons with $|\eta|<2.5$ and $p_{T} > 30~\mathrm{GeV}$. We examine events for jets using the inclusive Cambridge-Aachen jet algorithm with $R = 1.2$ and accept an event only if it has a jet with $P_T \ge 200\ \mathrm{GeV}$ and $|\eta|<2.5$. This is the same as the event selection in ref.~\cite{Butterworth:2008iy}.

Having generated events, we now analyze them to look for the $ZH$ signal. We will use the trimming and pruning analyses described in the previous section and, in addition, we will use the filtering method. Thus we need to describe the filtering method \cite{Butterworth:2008iy}, which has been applied several times in association with Higgs searches \cite{Plehn:2009rk,Kribs:2009yh}.
  
To use filtering, we first look for jets in the event using the Cambridge-Aachen (C-A) algorithm with $R = 1.2$ and select the highest $P_T$ jet, the ``fat jet.'' Then we examine the fat jet for a mass drop. If we have a signal event, then one of the splittings in the C-A splitting history is likely to be the $H \to b \bar b$ splitting. To look for it, we start at the trunk of the splitting tree and look at the first splitting, $J_{\{ij\}} \to J_i + J_j$. If the jet mass change in this splitting is large enough,
\begin{equation}
\max(M_i,M_j) < \mu M_{\{ij\}}
\end{equation}
with $\mu = 0.67$, and if the transverse momentum in the splitting is large enough,
\begin{equation}
\frac{\min(P_{T,i}^2,P_{T,j}^2)}{M_{\{ij\}}^2}
\left[(y_i - y_j)^2 + (\phi_i - \phi_j)^2 \right]
> y_{\rm cut}
\end{equation}
with $y_{\rm cut} = 0.09$, then we say that the mass drop condition is met and proceed to the next stage of the analysis. If the mass drop condition is not met, we eliminate the daughter jet with the smaller $P_T$ and examine the splitting of the daughter jet with the larger $P_T$ to see if its splitting satisfies the mass drop condition. This process continues until the mass drop condition is met.

It can be that the mass drop condition is never met. In this case the event is removed from the sample. This has the possibility of preferentially removing background events. 

If the mass drop condition is met, we apply a different analysis, called filtering, to the daughter protojets $i$ and $j$. First, we are hoping that this was a $H \to b \bar b$ splitting, so we ask whether both protojets $i$ and $j$ are tagged as containing a $b$ or $\bar b$ quark. For each of the two protojets, we assume a $b$-tagging efficiency of 60\% and a mistagging probability of 2\%. If one or both protojets $i$ and $j$ are not tagged as $b$-jets, the event is removed from the analysis. If both protojets $i$ and $j$ are tagged as $b$-jets, we apply the C-A algorithm separately to the constituents of both of these protojets. This time, we use a smaller cone size
\begin{equation}
R = \min\left(\frac{1}{2}\, 
\left[(y_i - y_j)^2 + (\phi_i - \phi_j)^2 \right]^{1/2}, 0.3\right)
\;\;.
\end{equation}
This procedure simply combines the branches of the C-A splitting trees for protojets $i$ and $j$ down to the level specified by this $R$.

We arrive at a list of constituent subjets of jets $i$ and $j$. We hope that the two highest $P_T$ subjets thus found each contain one of the previously found $b$ or $\bar b$ quarks, so we ask if they do. If we have a double $b$-tag in this sense, we retain the event, otherwise we remove it from the event sample. 

We next look at the three highest $P_T$ jets among the subjets of $i$ and $j$. The constituents of these constitute the filtered jet. The final step of the filtering analysis is to measure the mass $M_{\rm Jet}^{(f)}$ of the filtered jet.

Having explained filtering, we are now ready to compare filtering, trimming, and pruning and combinations of them. Here, we will be somewhat tethered to filtering even if we do not use the filtered jet mass $M_{\rm Jet}^{(f)}$. That is because it has been shown \cite{btags} that the filtering method described above is an efficient way for selecting events with double $b$-tags. Therefore, we use the filtering method to reject events that do not have double $b$-tags. In the $b$-tagged sample, we can measure any of the filtered jet mass, $M_{\rm Jet}^{(f)}$, the trimmed jet mass, $M_{\rm Jet}^{(t)}$, or the pruned jet mass, $M_{\rm Jet}^{(p)}$. We apply the trimming or pruning procedures independently of any subjet information from the mass drop plus filtering procedure. We know that the event was selected because there was a suitable mass drop and the appropriate subjets were $b$-tagged, but we do not ask that the subjets used in trimming or pruning have any particular relation to the subjets that had been found in the mass drop plus filtering procedure. For trimming, we follow the method outlined in section \ref{sec:top}, defining the fat jet using the anti-$k_T$ algorithm, but with $R = 1.2$ instead of the value $R = 1.5$ used in section \ref{sec:top}. For pruning, we follow the method outlined in section \ref{sec:top}, defining the fat jet using the C-A algorithm, again with $R = 1.2$. For filtering, we follow the method outlined in this section.

\FIGURE{
\includegraphics[width=7.0cm]{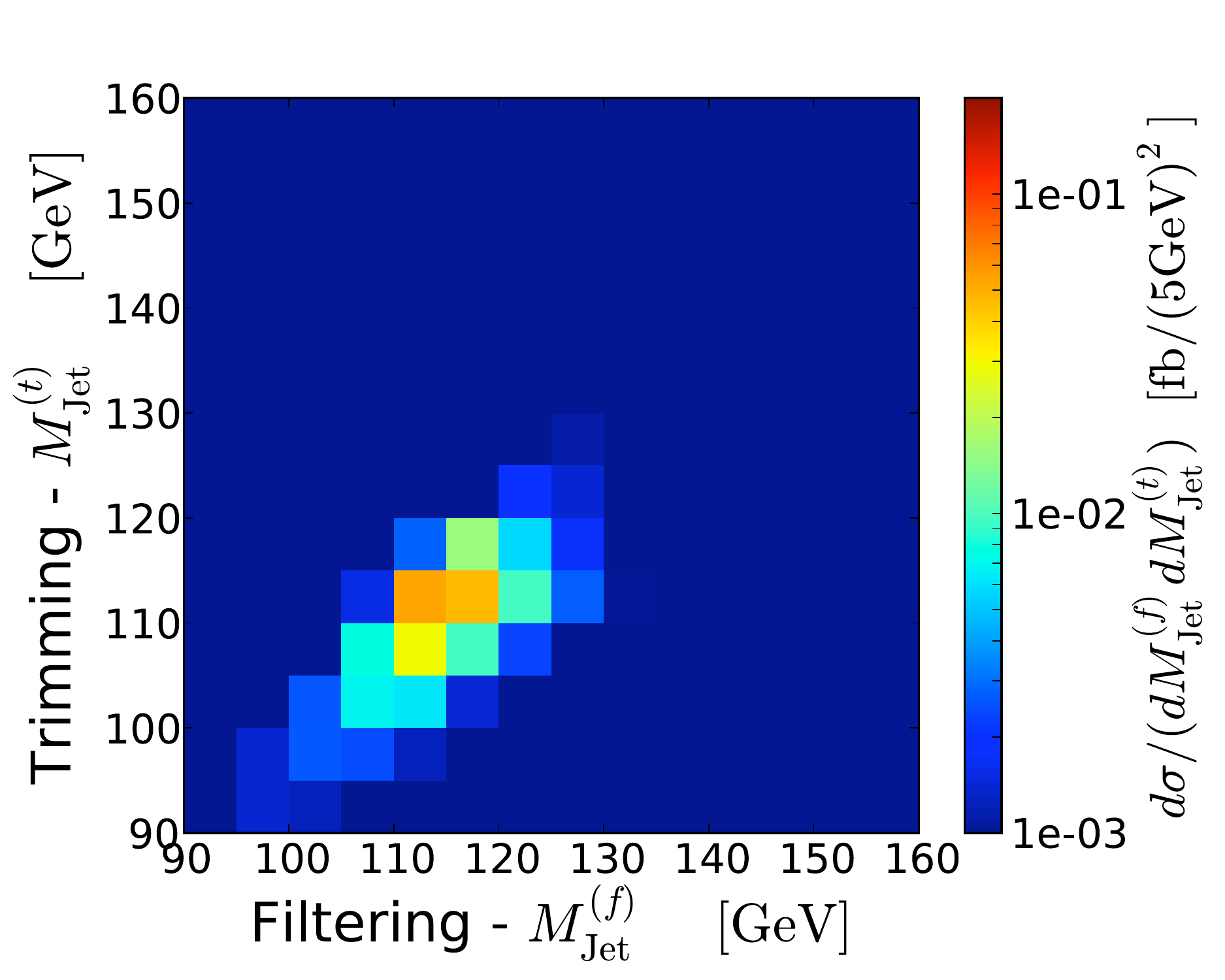} 
\includegraphics[width=7.0cm]{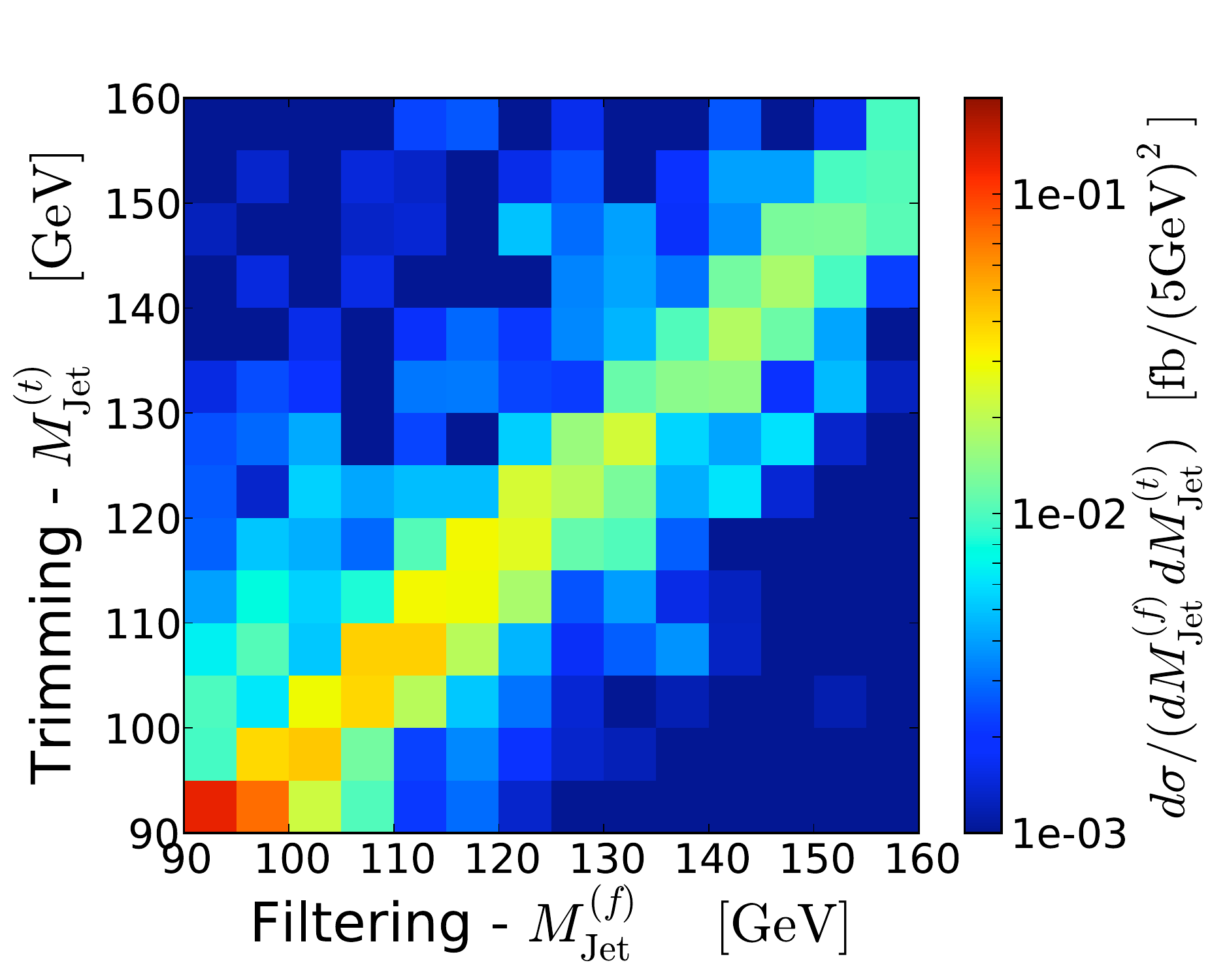} \\
\includegraphics[width=7.0cm]{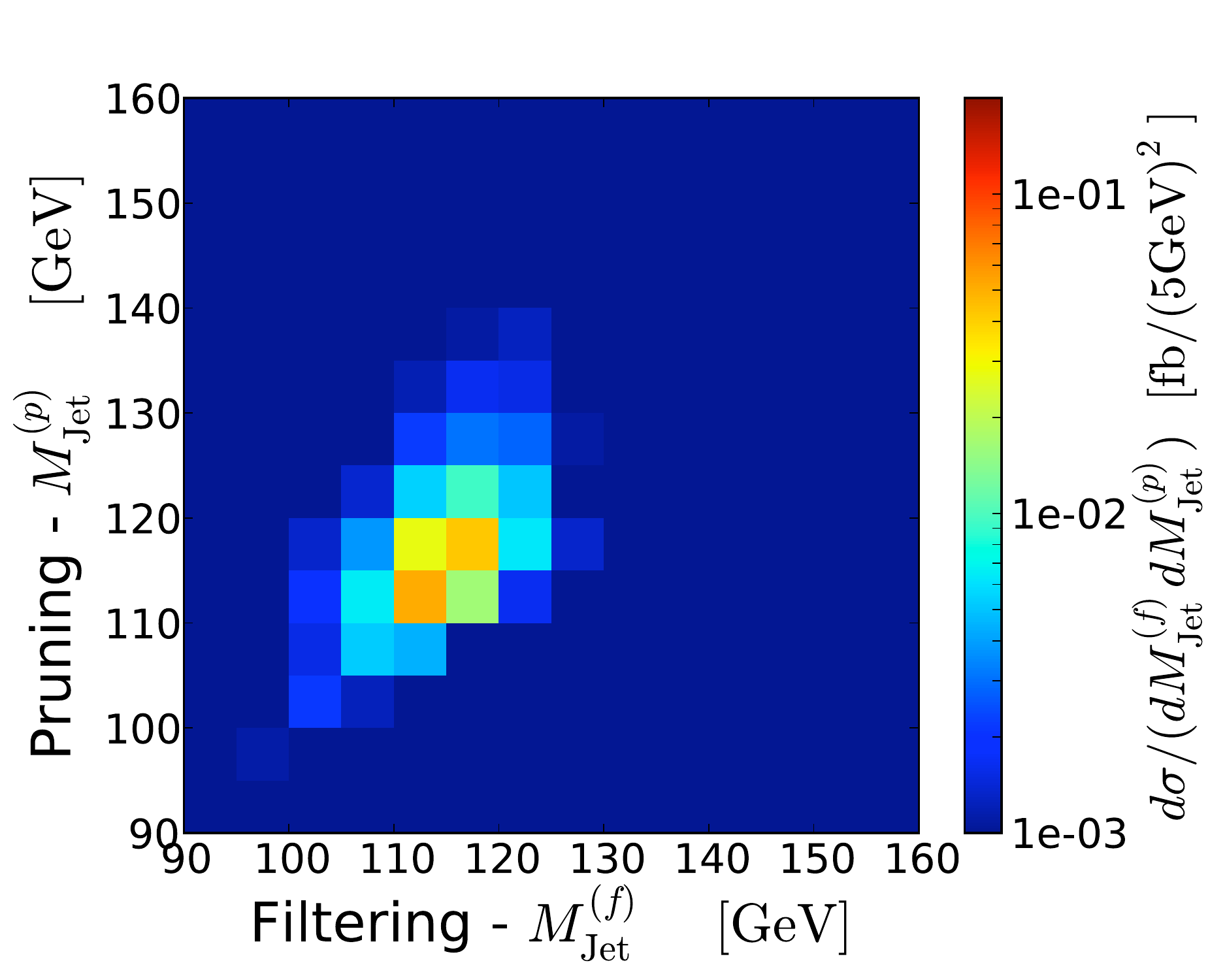}
\includegraphics[width=7.0cm]{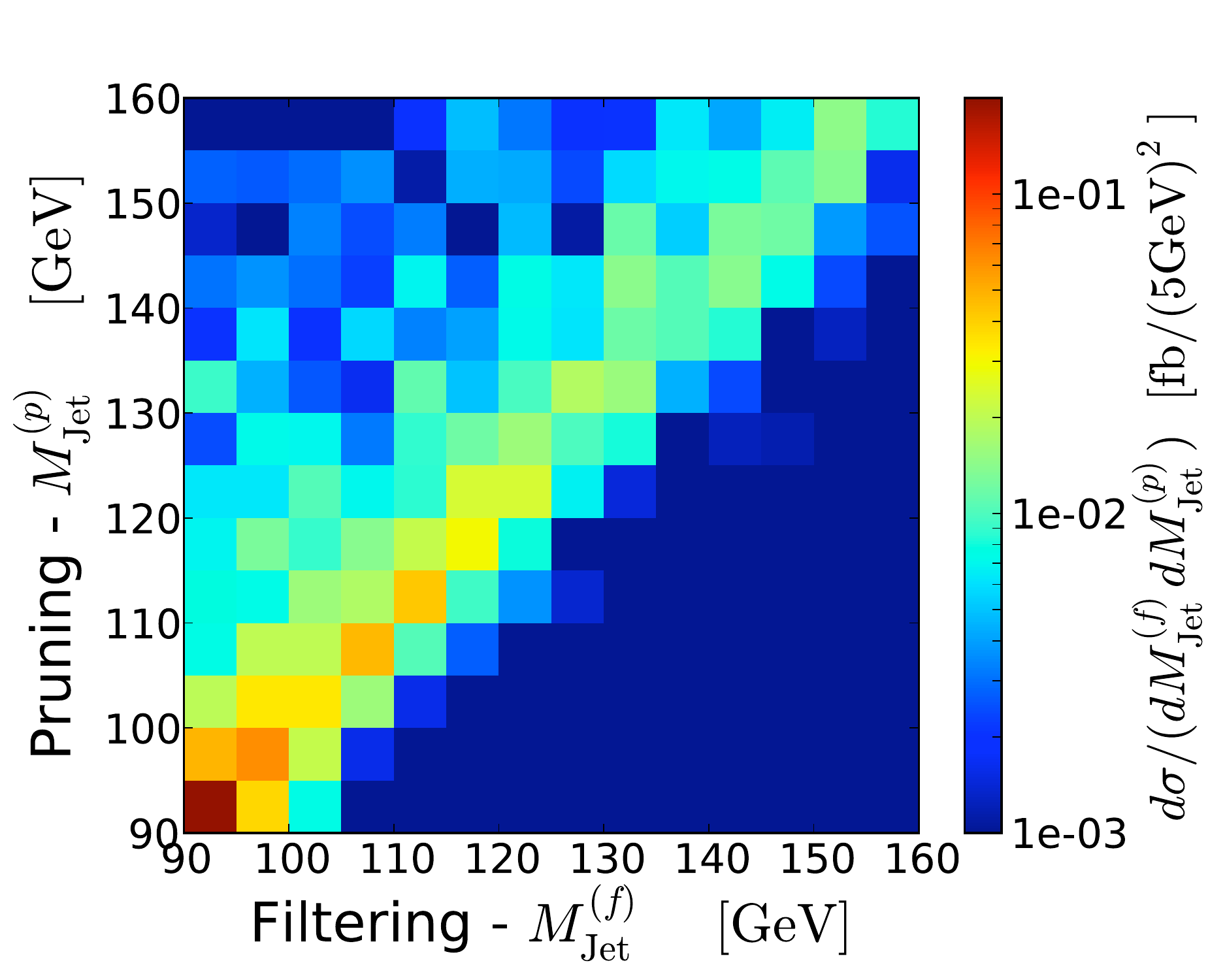} \\
\includegraphics[width=7.0cm]{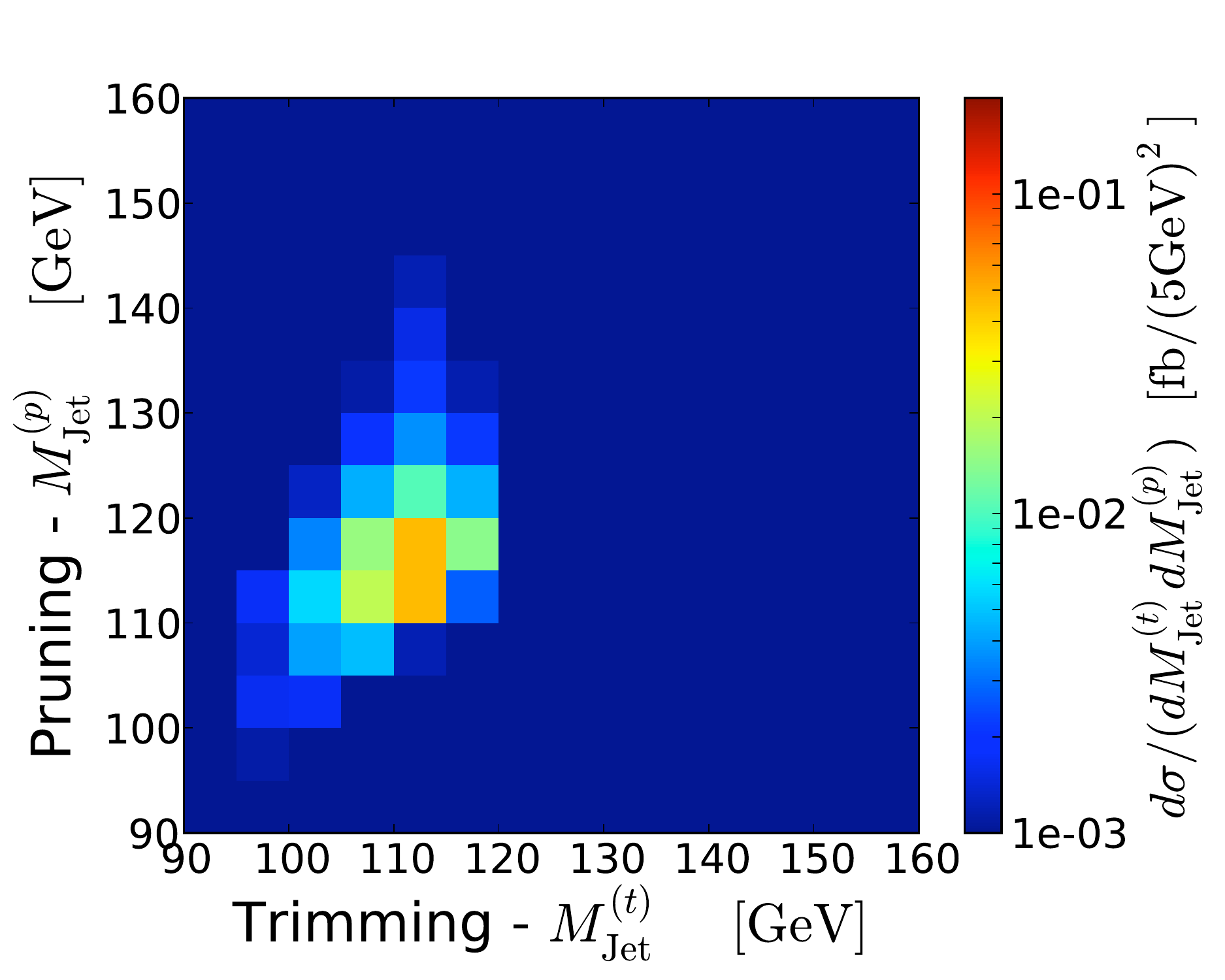} 
\includegraphics[width=7.0cm]{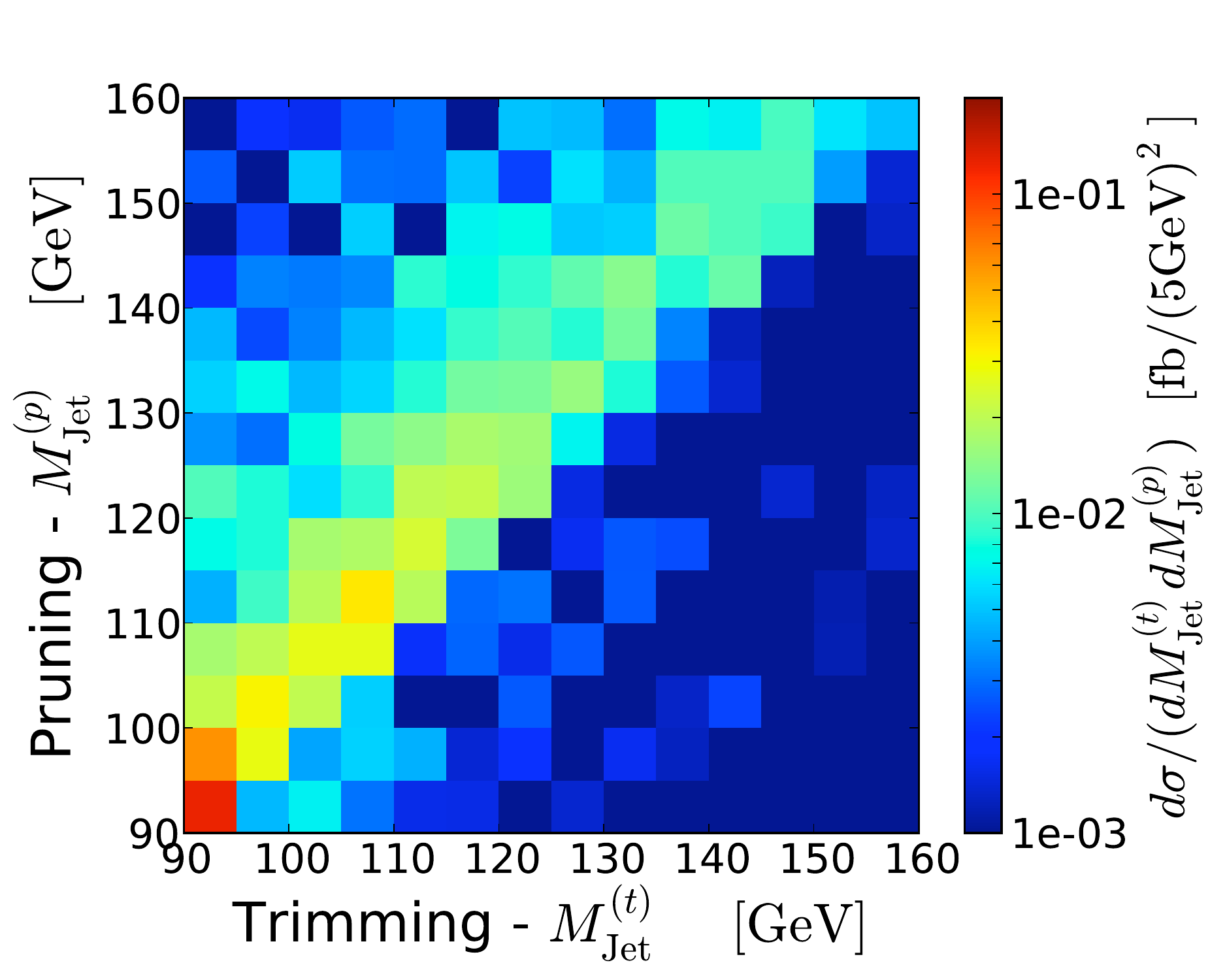} 
\caption{Joint distributions between pairs of the filtered jet mass $M_{\rm Jet}^{(f)}$, the trimmed jet mass $M_{\rm Jet}^{(t)}$, and the pruned jet mass $M_{\rm Jet}^{(p)}$ for the $ZH$ signal (left column) and the background (right column). The events were generated with $M_{\rm Higgs} = 115\ {\rm GeV}$.
}
\label{fig:ZHmass2D}
}

We note that in a given event, the three mass measures, $M_{\rm Jet}^{(f)}$, $M_{\rm Jet}^{(t)}$, and $M_{\rm Jet}^{(p)}$ are not the same. We illustrate this in figure~\ref{fig:ZHmass2D}, in which we plot the numbers of events in bins of $(M_{\rm Jet}^{(t)},M_{\rm Jet}^{(f)})$ (top row), $(M_{\rm Jet}^{(p)},M_{\rm Jet}^{(f)})$ (middle row), and $(M_{\rm Jet}^{(p)},M_{\rm Jet}^{(t)})$ (bottom row). These distributions are shown for the $ZH$ signal in the left hand column and for the background in the right hand column. In each graph, we see that knowing one of the two masses does not tell us the other. There is, of course, a correlation, but it is far from perfect. For that reason, there is information to be gained by measuring two of these masses for each event. (We do not have enough generated events to divide them into a three dimensional grid of masses.)

With this in mind, we choose a mass window for each of $M_{\rm Jet}^{(f)}$, $M_{\rm Jet}^{(t)}$, and $M_{\rm Jet}^{(p)}$, namely $W_f = (110\ {\rm GeV}, 125\ {\rm GeV})$, $W_t = (105\ {\rm GeV}, 120\ {\rm GeV})$, and $W_p = (110\ {\rm GeV}, 125\ {\rm GeV})$. Then, to start, we assume an integrated luminosity $\int\!dL = 30\ {\rm fb}^{-1}$ is available and we count the number of signal events $s$ and background events $b$ expected with the mass of the filtered jet in its window, $M_{\rm Jet}^{(f)} \in W_{f}$. The results are displayed in the first column of table \ref{tab:MHinbins}. Using just this information, the expected logarithm of the likelihood ratio favoring the presence of the $ZH$ signal along with the background, eq.~(\ref{eq:expectedA}), is $\langle {\cal L}(n)\rangle_{\rm SB} \approx 1.7$. Then we ask that both $M_{\rm Jet}^{(f)}$ and $M_{\rm Jet}^{(t)}$ be in their respective mass windows. This cuts the number of signal and background events, but makes $\langle {\cal L}(n)\rangle_{\rm SB}$ larger, indicating a greater statistical significance for the measurement. Similarly, we find larger values of $\langle {\cal L}(n)\rangle_{\rm SB}$ both for the combination of  $M_{\rm Jet}^{(f)}$ and $M_{\rm Jet}^{(p)}$ and for the combination of $M_{\rm Jet}^{(p)}$ and $M_{\rm Jet}^{(t)}$. We point out here that we are looking only at statistical significance from counting statistics, not at other sources of error. Additionally, we note that $\langle {\cal L}(n)\rangle_{\rm SB} \approx 2$ is not nearly enough to claim a discovery of the signal; however, if one had $\langle {\cal L}(n)\rangle_{\rm SB} \approx 4$ from another independent method, such as a search for $WH$ production, then the ability to add 2 to this would be not insignificant.

\TABLE{
\begin{tabular}{c|c|c|c|c}
  &  
$\begin{array}{c} M_{\rm Jet}^{(f)} \in W_{f} \\  \end{array}$       & 
$\begin{array}{c} M_{\rm Jet}^{(f)} \in W_{f} \\ 
                  M_{\rm Jet}^{(t)} \in W_{t} \end{array}$ & 
$\begin{array}{c} M_{\rm Jet}^{(f)} \in W_{f} \\ 
                  M_{\rm Jet}^{(p)} \in W_{p} \end{array}$ &  
$\begin{array}{c} M_{\rm Jet}^{(p)} \in W_{p} \\ 
                  M_{\rm Jet}^{(t)} \in W_{t} \end{array}$   \\
	\hline 
Signal cross section [fb]     & 0.20 & 0.18 & 0.17 & 0.17 \\
Backgrnd cross section [fb]\!   & 0.30 & 0.20 & 0.17 & 0.16 \\ 
$s/b$                         & 0.67 & 0.90 & 1.0  & 1.1 \\
$s/\sqrt{b}$ \ \ ($\int\!dL = 30\ {\rm fb}^{-1}$) 
                              & 2.0 & 2.2 & 2.3 & 2.3 \\
\!$\langle{\cal L}(n)\rangle_{\rm SB}$  ($\int\!dL = 30\ {\rm fb}^{-1}$)\!\!
                              & 1.7 & 1.9 & 2.0 & 2.1
\end{tabular}
\label{tab:MHinbins}
\caption{Statistical significance of $ZH$ results for an integrated luminosity  of $30\ {\rm fb}^{-1}$. Here we simply count the expected number of signal events, $s$, and background events, $b$, in certain windows for the mass of the filtered jet, $M_{\rm Jet}^{(f)}$, the mass of the trimmed jet, $M_{\rm Jet}^{(t)}$, and the mass of the pruned jet, $M_{\rm Jet}^{(p)}$. The mass windows chosen are $W_f = (110\ {\rm GeV}, 125\ {\rm GeV})$, $W_t = (105\ {\rm GeV}, 120\ {\rm GeV})$, and $W_p = (110\ {\rm GeV}, 125\ {\rm GeV})$. The Higgs mass assumed when generating events is $M_{\rm Higgs} = 115\ {\rm GeV}$. In the first column, we ask only that the filtered jet mass be in the window $W_f$. In the remaining columns, we combine methods by asking that two masses be in the corresponding windows. For each type of measurement, we show three measures of statistical significance, $s/b$,  $s/\sqrt{b}$, and the logarithm of the likelihood ratio based on $s$ and $b$, eq.~(\ref{eq:expectedA}).
}
}

It is rather limiting to base the assessment of whether data favors the presence of the $ZH$ signal in addition to the background on simply the counts in a single window in a pair of jet masses. As we noted in our example of $t\bar t$ production in section \ref{sec:top}, the experiment will give counts $n_J$ in each bin $J$ shown in figure \ref{fig:ZHmass2D}. Again, we can base our assessment on the log likelihood ratio using all of the information.\footnote{We include all of the bins that contain a background cross section of at least $0.001\ {\rm fb}$. The results are not sensitive to this cut, which we impose so that we can have a reliable calculation of $s_J/b_J$.} Then the likelihood ratio is the product of the likelihood ratios for all of the bins used. Its logarithm is given by
\begin{equation}
\label{eq:LfullwithmassH}
{\cal L}(\{n\},m) = 
\sum_J 
\left[n_J 
\log\left(1 + \frac{s_J(m)}{b_J}\right) - s_J(m)\right]
\;\;.
\end{equation}
Here we have included in the notation the fact that the expected number of signal events $s_J$ in a certain bin depends on the assumed Higgs boson mass, $m$. Given data $\{n\}$, one can test not only whether the presence of the $ZH$ signal is favored, but how the likelihood favoring the presence of the signal depends on the assumed mass $m$. The expectation value of ${\cal L}(\{n\},m)$ if the true Higgs boson mass is $M_{\rm Higgs}$ and the signal is present along with the background is given by
\begin{equation}
\label{eq:expectedLfullwithmassH}
\langle {\cal L}(\{n\},m)\rangle_{\rm SB} = 
\sum_J 
\left[(s_J(M_{\rm Higgs}) + b_J) 
\log\left(1 + \frac{s_J(m)}{b_J}\right) - s_J(m)\right]
\;\;.
\end{equation}
We have computed $\langle {\cal L}(\{n\},m)\rangle_{\rm SB}$ for nine assumed values of $m$ and for the three combinations of using two out of three of the filtered jet mass, the trimmed jet mass, and the pruned jet mass. The results are displayed in figure~\ref{fig:ZHplotsforL}.

\FIGURE{
\includegraphics[width=7.0cm]{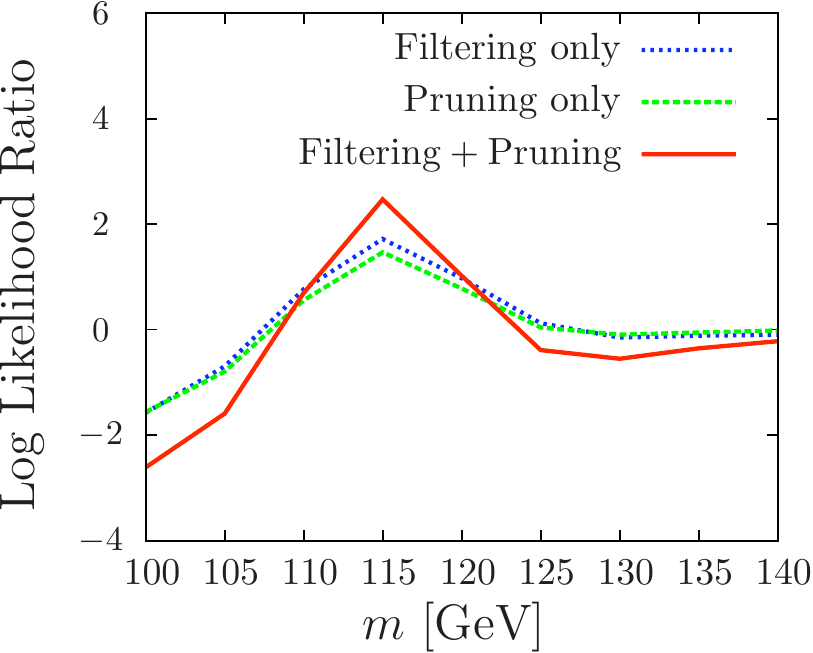}
\includegraphics[width=7.0cm]{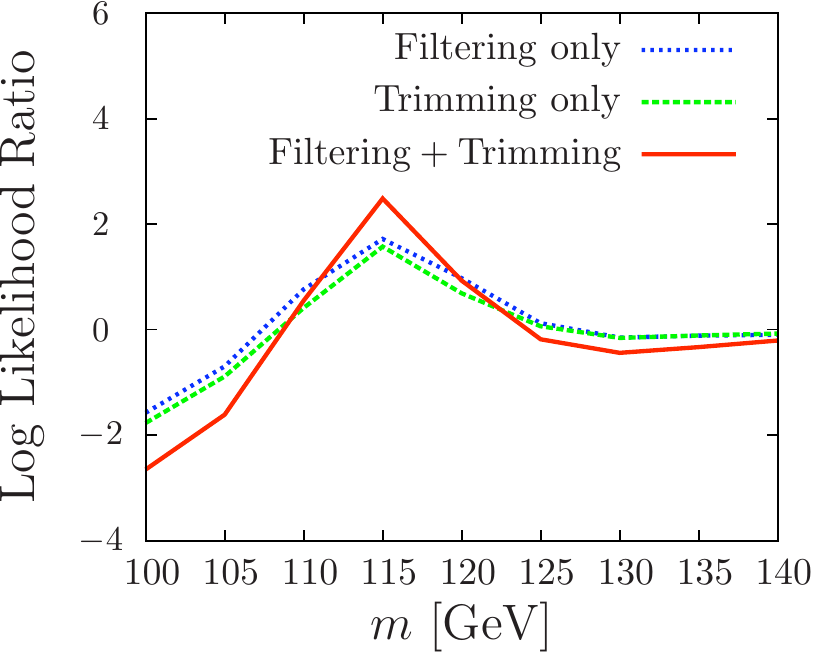} \\
\includegraphics[width=7.0cm]{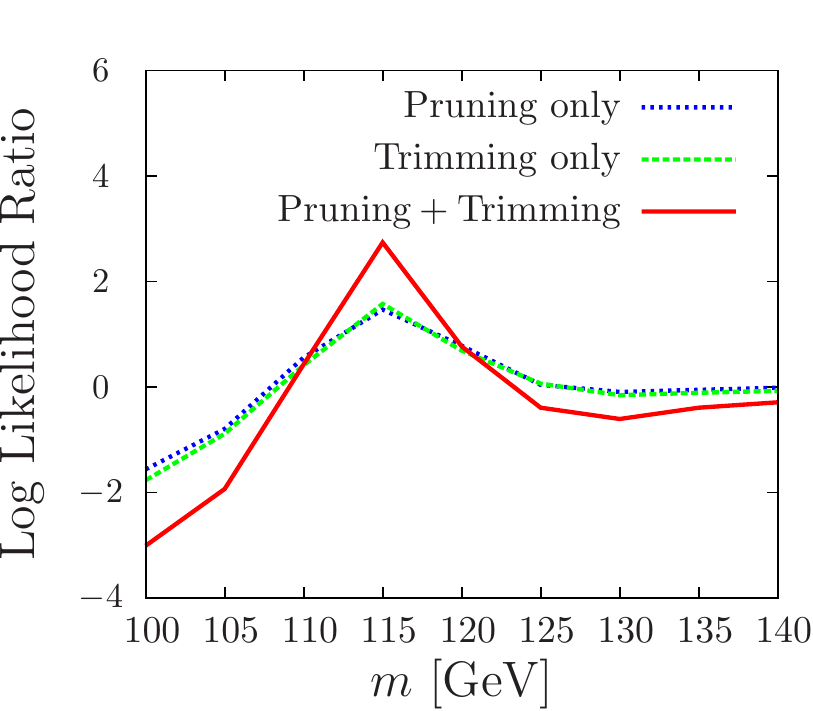}
\caption{The log likelihood ratio in the $ZH$ production process as a function of the trial Higgs boson mass $m$, assuming an integrated luminosity of $30\ {\rm fb}^{-1}$. We construct ${\cal L}(\{n\},m)$ for nine different values of $m$. Then we take the expectation value of these quantities in the signal + background theory with the true Higgs mass, $M_{\rm Higgs}$. The results are shown using the the filtered jet mass $M_{\rm Jet}^{(f)}$ alone, the trimmed jet mass $M_{\rm Jet}^{(t)}$ alone, the pruned jet mass $M_{\rm Jet}^{(p)}$ alone, and for $M_{\rm Jet}^{(f)}$ combined with $M_{\rm Jet}^{(p)}$, for $M_{\rm Jet}^{(f)}$ combined with $M_{\rm Jet}^{(t)}$ and for $M_{\rm Jet}^{(p)}$ combined with $M_{\rm Jet}^{(t)}$.
}
\label{fig:ZHplotsforL}
}

We learn three things from figure~\ref{fig:ZHplotsforL}. First, if we look at the case $m = M_{\rm Higgs}$, we have a stronger signal using the distribution in two out of the three variables $M_{\rm Jet}^{(f)}$, $M_{\rm Jet}^{(t)}$, and $M_{\rm Jet}^{(p)}$ together than we have for just one variable. Second, with the distribution in two variables, we have better resolution in which trial mass $m$ best fits the data compared to the resolution obtained with just one variable. Finally, using $M_{\rm Jet}^{(t)}$ together with $M_{\rm Jet}^{(p)}$, we have $\langle {\cal L}(\{n\},m)\rangle_{\rm SB} \approx 2.7$. This is better than the corresponding result, $\langle {\cal L}(\{n\},m)\rangle_{\rm SB} \approx 2.1$, from table \ref{tab:MHinbins}, in which we used simply the number of counts in a fixed window in $M_{\rm Jet}^{(t)}$ and $M_{\rm Jet}^{(p)}$.

We have tried one more small adjustment. In the pruning method, Refs.~\cite{Ellis:2009su,Ellis:2009me} recommend that the parameter $z_{\rm cut}$, Eq.~(\ref{eq:zcutdef}), be set to 0.1. That is the value we have used. However, we find that the value 0.05 does a better job in this application, as shown in table \ref{tab:MHinbinsnewz}. Changing to $z_{\rm{cut}}=0.05$ allows the pruned jet to absorb more soft radiation. This enhances the asymmetry in the jet mass between pruning and trimming. Although the correlation of the jet mass for the signal process is weakened it mainly affects the background of light parton jets, see figure~\ref{fig:ZHmass2DZ005}.

\TABLE{
\begin{tabular}{c|c}
  &    
$\begin{array}{c} M_{\rm Jet}^{(p)} \in W_{p} \\ 
                  M_{\rm Jet}^{(t)} \in W_{t} \end{array}$   \\
	\hline 
Signal cross section [fb]     &  0.16 \\
Backgrnd cross section [fb]   &  0.13 \\ 
$s/b$                         &  1.3 \\
$s/\sqrt{b}$ \ \ ($\int\!dL = 30\ {\rm fb}^{-1}$) 
                              &  2.4 \\
\hbox{\hskip 0.8 cm $\langle{\cal L}(n)\rangle_{\rm SB}$  ($\int\!dL = 30\ {\rm fb}^{-1}$)\hskip 0.8 cm }
                              &  2.2
\end{tabular}
\label{tab:MHinbinsnewz}
\caption{Statistical significance of $ZH$ results for an integrated luminosity of $30\ {\rm fb}^{-1}$ as in table \protect\ref{tab:MHinbins} except that here we take $z_{\rm cut}$ in the pruning method to be 0.05 instead of 0.1. This improves the statistical significance compared to the $(M_{\rm Jet}^{(p)},M_{\rm Jet}^{(t)})$ results in the rightmost column of table \protect\ref{tab:MHinbins}.
}
}

\FIGURE{
\includegraphics[width=7.0cm]{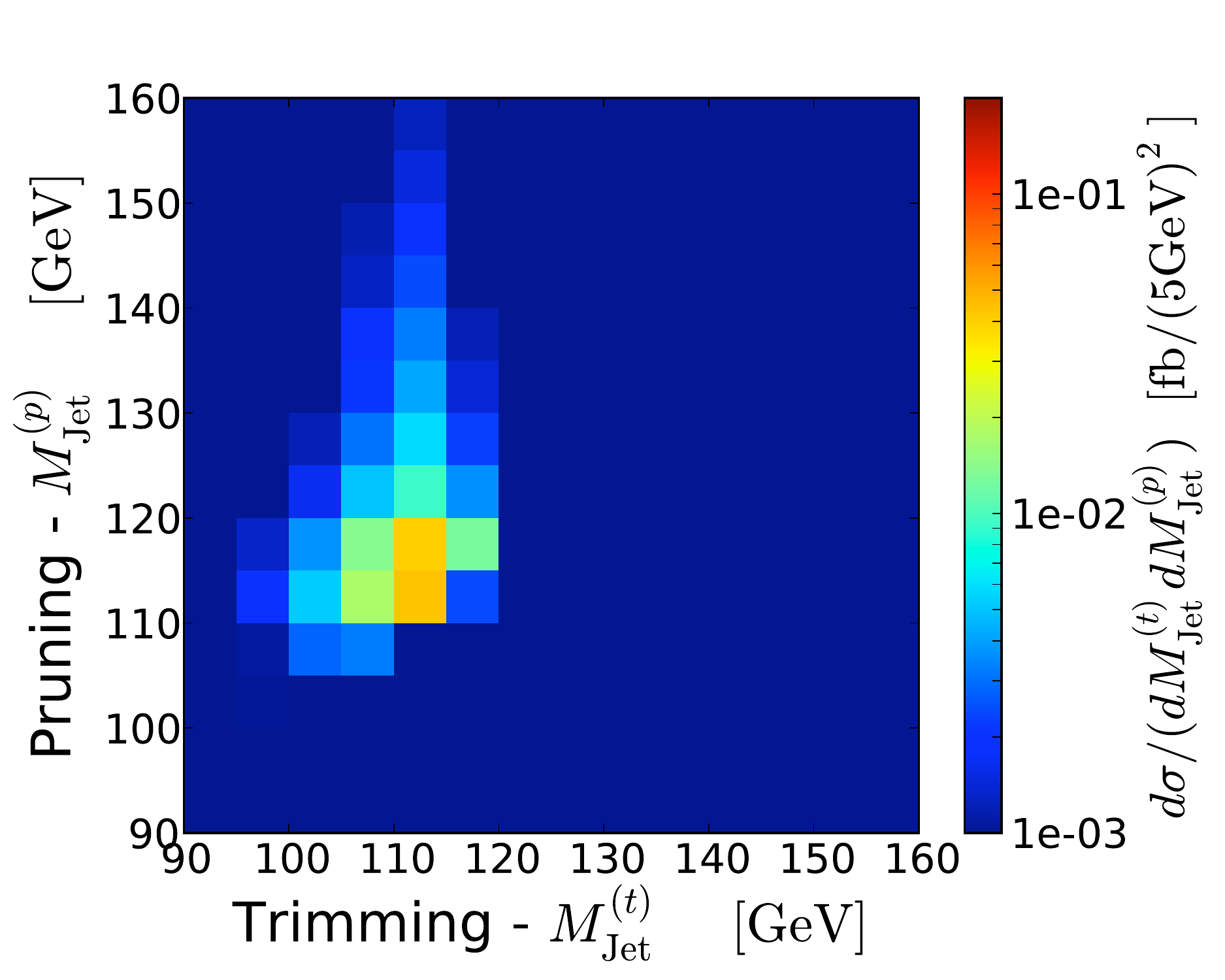} 
\includegraphics[width=7.0cm]{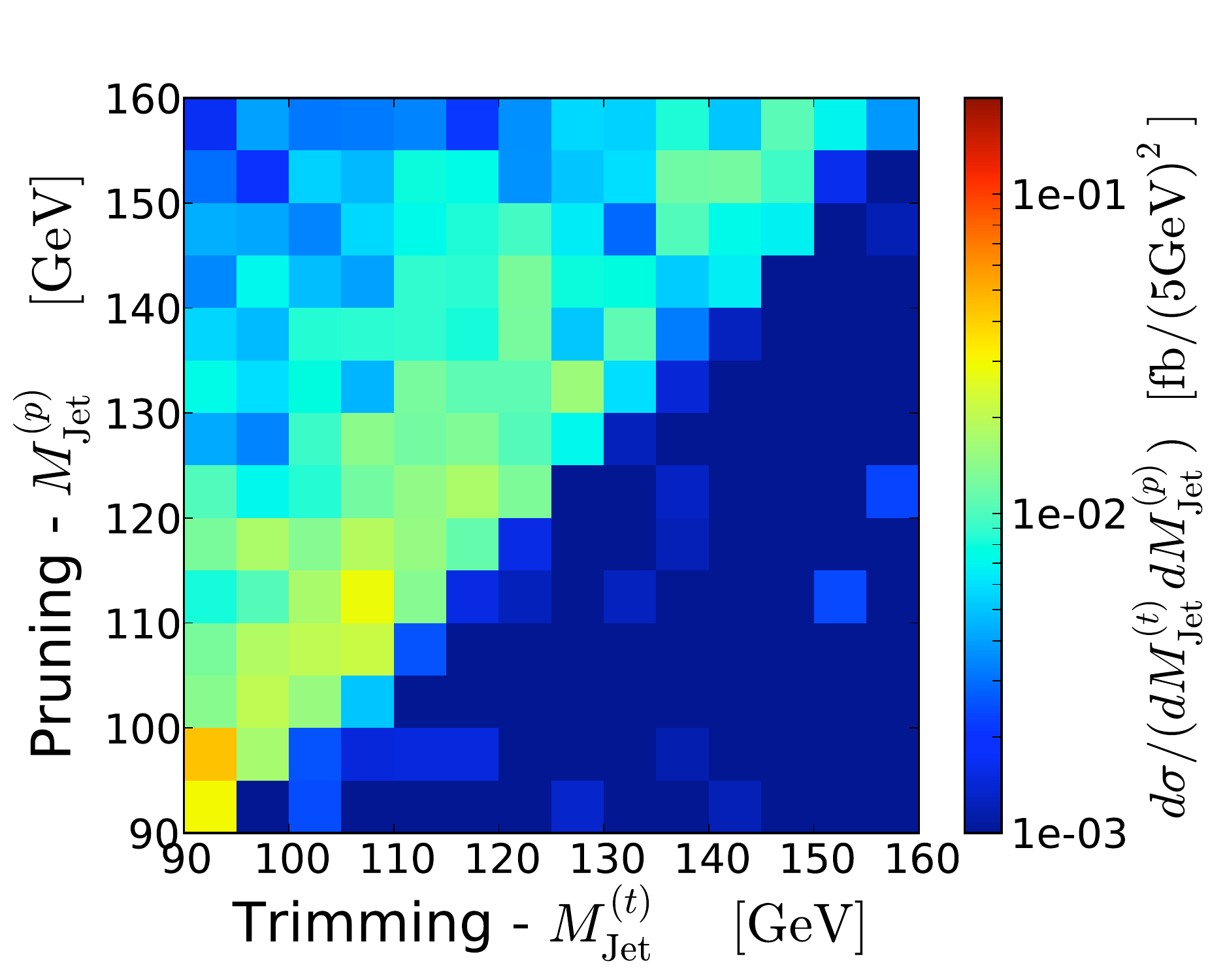} 
\caption{Joint distributions between pairs of the trimmed jet mass $M_{\rm Jet}^{(t)}$ and the pruned jet mass $M_{\rm Jet}^{(p)}$ for the $ZH$ signal (left column) and the background (right column). The events were generated with $M_{\rm Higgs} = 115\ {\rm GeV}$ and $z_{\rm{cut}}=0.05$.
}
\label{fig:ZHmass2DZ005}
}

One can well be concerned that smearing of jet masses because of detector effects might affect the results presented here. To check, we applied Gaussian smearing on $M_{\rm Jet}^{(f)}$, $M_{\rm Jet}^{(t)}$ and $M_{\rm Jet}^{(p)}$ according to \cite{smearing} but could not find sizable differences in the log likelihood ratio. More realistic finite jet resolution effects might change the quantitative statements in this paper, but a reliable simulation of them is beyond the scope of our work. Thus all results shown are without detector smearing effects.

\section{Conclusion}
\label{sec:conclusion}

In searches for a narrow boosted resonance in which the signal is small compared to a background coming from QCD induced light parton jets, the combination of pruning, trimming and filtering can help to extract the signal from the background. 

For a generic resonance tagger, one can look for excess events in a combined window of two of the pruned, trimmed, and filtered jet masses. Even if a new physics model for the signal is not anticipated, the approach outlined in this paper can be used to improve on the statistical significance of a so-called ``side bin analysis.'' Both $s/b$ and $s/\sqrt b$ can be improved for a given window around the resonance mass when different jet mass measures are used together. 

If one has good models for the expected background and the sought signal, one can gain further statistical significance by using a likelihood analysis based on the models for signal and background.

\acknowledgments

We thank Chris Vermilion and David Krohn for helpful comments on the pruning and trimming procedures and for carefully reading a draft of this paper and we thank Giacinto Piacquadio, Gavin Salam, and David Reeb for helpful discussions. This work originated at the Workshop on Jet Substructure in the series Northwest Workshops on Terascale Physics. The workshop was held at the University of Washington under DOE grant DE-FG02-96ER40956. We thank Steve Ellis and Ann Nelson for arranging the workshop. This work was supported by DOE grant DE-FG02-96ER40969.

\appendix

\section{Analysis of data with relative likelihoods}  \label{sec:Likely}

In this paper, we have made use of analysis of data using relative likelihoods. This method, in one form or another, is quite widely used. A description convenient for our use can be found in Ref.~\cite{GU:Stat}. We provide a brief summary in this appendix.

We suppose that we measure one or several variables $\vec v = (v_1, v_2, \dots, v_L)$ for each event. We put the events into bins, with labels $J$, based on the value of $\vec v$ of each event. Let $n_J$ be the number of events in bin $J$. Then the result of the experiment is a list of the values $\{n\} = \{n_1,\cdots,n_N\}$ of the numbers of events in each bin.

We suppose that we have a model (say, based on \textsc{Pythia}) for the expected number of events in each bin if there is no new physics signal. This is the background model, designated B. Let us denote the expectation value of $n_J$ in the background model by $b_J$. We also suppose that we have a model for the expected number of events in each bin if there is a certain new physics signal. This is the model then includes both the background and the sought signal. We call this model SB. Let us denote the expectation value of $n_J$ in the signal plus background model as $b_J + s_J$. For the moment, we assume that there is no uncertainty in what models B and SB predict.

Given model B, the probability to find result $\{n\}$ is
\begin{equation}
P_{\rm B}(\{n\}) = \prod_J \frac{1}{n_J!}\, (b_J)^{n_J} e^{-b_J}
\;\;.
\end{equation}
Given model SB, the probability to find result $\{n\}$ is
\begin{equation}
P_{\rm SB}(\{n\}) = \prod_J \frac{1}{n_J!}\, (b_J + s_J)^{n_J} e^{-b_J-s_J}
\;\;.
\end{equation}
The ratio of these is the relative likelihood to find the observed result,
\begin{equation}
R(\{n\}) = \frac{P_{\rm SB}(\{n\})}{P_{\rm B}(\{n\})}
\;\;.
\end{equation}
This ratio tells one how to modify a prior opinion about the probability that SB as opposed to B holds in nature. Thus it is a convenient statistic to describe the results of the experiment.

We can write the likelihood ratio as
\begin{equation}
R(\{n\}) = \exp{{\cal L}(\{n\})}
\;\;.
\end{equation}
The logarithm of the likelihood ratio, ${\cal L}(\{n\})$, has a simple expression
\begin{equation}
{\cal L}(\{n\}) = \sum_J [n_J \log(1 + s_J/b_J) - s_J]
\;\;.
\end{equation}

The theory for signal and background can depend on parameters, so that $s_J$ and $b_J$ depend on the parameters. Then ${\cal L}(\{n\})$ depends on the parameters. Given data $\{n\}$, we can adjust the parameters to find the version of the theory with the biggest ${\cal L}(\{n\})$. In this paper, we consider the simple case in which there is a single parameter\footnote{If $s_J/b_J \ll 1$ in the bins with the most signal, then it is important to know the normalization of the background quite precisely. In this case, one might introduce a parameter $\lambda$ that represents the normalization of the background and use the data to fix $\lambda$.} that we consider varying, a mass that we denote by $m$. The signal depends on $m$; the background does not. Thus
\begin{equation}
{\cal L}(\{n\},m) = \sum_J [n_J \log(1 + s_J(m)/b_J) - s_J(m)]
\;\;.
\end{equation}

To understand this better, it is useful to consider the case in which $s_J \ll b_J$ and $(n_J - b_J) \ll b_J$ in all bins. Then
\begin{equation}
\begin{split}
{\cal L} ={}& \sum_J \left\{
[b_J + (n_J - b_J)] 
\left[\frac{s_J(m)}{b_J} - \frac{s_J(m)^2}{2 b_J^2}+\cdots\right] - s_J(m)
\right\}
\\
={}&
\sum_J \left\{
s_J(m)
+(n_J - b_J)\,\frac{s_J(m)}{b_J}
- \frac{s_J(m)^2}{2 b_J}+\cdots
- s_J(m)
\right\}
\\
\approx{}&
\sum_J \left\{
\frac{(n_J - b_J)s_J(m)}{b_J}
- \frac{s_J(m)^2}{2 b_J}
\right\}
\;\;.
\end{split}
\end{equation}
This has a simple interpretation. We see that ${\cal L}$ is large when the observed signal $(n_J - b_J)$ is correlated with the expected signal $s_J(m)$. That is, ${\cal L}$ is large when $(n_J - b_J) > 0$ in those bins for which $s_J(m) > 0$. There is a penalty contribution, ${s_J(m)^2}/{(2 b_J)}$ for each bin. Thus, to keep ${\cal L}$ positive, $(n_J - b_J)$ needs to be bigger than $s_J(m)/2$ in the bins with expected signal.

Suppose that the SB theory is correct if we set the mass to its true value $M_{\rm true}$. The expected value of ${\cal L}(\{n\},m)$ in this case is
\begin{equation}
\langle{\cal L}\rangle = \sum_J [\bar n_J \log(1 + s_J/b_J) - s_J]
\;\;,
\end{equation}
where
\begin{equation}
\bar n_J = b_j + s_J(M_{\rm true})
\;\;.
\end{equation}
In the case of small signal/background, we have
\begin{equation}
\langle{\cal L}\rangle \approx 
\sum_J \left\{
\frac{ s_J(M_{\rm true})s_J(m)}{b_J}
- \frac{s_J(m)^2}{2 b_J}
\right\}
\;\;.
\end{equation}
When we set $m$ to $M_{\rm true}$, this is
\begin{equation}
\langle{\cal L}(\{n\},M_{\rm true})\rangle \approx 
\sum_J 
\frac{s_J(M_{\rm true})^2}{2 b_J}
\;\;.
\end{equation}
That is, what counts in this limit is $s_J/\sqrt{b_J}$.

It is a simple matter to evaluate how the observed value of ${\cal L}(\{n\},m)$ fluctuates assuming that the SB theory is correct when $m = M_{\rm true}$. With Poisson statistics, we have
\begin{equation}
\begin{split}
\langle n_J \rangle ={}& \bar n_J
\;\;,
\\
\langle n_J^2 \rangle ={}& \bar n_J^2 + \bar n_J
\;\;.
\end{split}
\end{equation}
Let us adopt the shorthand notation
\begin{equation}
L_J = \log\left(1 + \frac{s_J(m)}{b_J}\right)
\;\;.
\end{equation}
Then
\begin{equation}
{\cal L} = \sum_J [n_J L_J - s_J]
\;\;.
\end{equation}
Let us denote the expectation value of ${\cal L}$ by $\overline{\cal L}$,
\begin{equation}
\overline{\cal L} = \sum_J [\bar n_J L_J - s_J]
\;\;.
\end{equation}
Then the variance of ${\cal L}$ is
\begin{equation}
\begin{split}
\big\langle ({\cal L} - \overline{\cal L})^2\big\rangle
={}& 
\left\langle \left(
\sum_J (n_J - \bar n_J) L_J
\right)^{\!\!2}\, \right\rangle
\\
={}& \sum_{J,K}L_J L_K
\big\langle 
(n_J - \bar n_J) (n_K - \bar n_K)\big\rangle
\\
={}& \sum_{J} \bar n_J L_J^2
\;\;.
\end{split}
\end{equation}
%



\begin{thebibliography}{99}

\bibitem{Butterworth:2008iy}
  J.~M.~Butterworth, A.~R.~Davison, M.~Rubin and G.~P.~Salam,
  {\em Jet substructure as a new Higgs search channel at the LHC},
  Phys.\ Rev.\ Lett.\  {\bf 100}, 242001 (2008) 
  [arXiv:0802.2470]
  \href{http://www.slac.stanford.edu/spires/find/hep/www?j=PRLTA,100,242001}
  {[SPIRES]}.

\bibitem{Krohn:2009th}
  D.~Krohn, J.~Thaler and L.~T.~Wang,
  {\em Jet Trimming},
  JHEP {\bf 1002}, 084 (2010)
  [arXiv:0912.1342]
  \href{http://www.slac.stanford.edu/spires/find/hep/www?j=JHEPA,1002,084}
  {[SPIRES]}.

\bibitem{Ellis:2009su}
  S.~D.~Ellis, C.~K.~Vermilion and J.~R.~Walsh,
  {\em Techniques for improved heavy particle searches with jet substructure},
  Phys.\ Rev.\  D {\bf 80}, 051501 (2009) 
 [arXiv:0903.5081]
 \href{http://www.slac.stanford.edu/spires/find/hep/www?j=PHRVA,D80,051501}
  {[SPIRES]}.
  
\bibitem{Ellis:2009me}
  S.~D.~Ellis, C.~K.~Vermilion and J.~R.~Walsh,
  {\em Recombination Algorithms and Jet Substructure: Pruning as a Tool for Heavy
   Particle Searches},
  arXiv:0912.0033 [hep-ph]
  \href{http://www.slac.stanford.edu/spires/find/hep/www?bb=ARXIV:0912.0033}
  {[SPIRES]}.

\bibitem{Brooijmans:2008zz}
  G.~Brooijmans,
  {\em High $p^T$ hadronic top quark identification. Part I: Jet mass and
  Ysplitter}, 
  ATL-COM-PHYS-2008-001
  \href{http://www.slac.stanford.edu/spires/find/hep/www?r=ATL-COM-PHYS-2008-001}
  {[SPIRES]}.
  
\bibitem{Kaplan:2008ie}
  D.~E.~Kaplan, K.~Rehermann, M.~D.~Schwartz and B.~Tweedie,
  {\em Top Tagging: A Method for Identifying Boosted Hadronically Decaying Top
  Quarks},
  Phys.\ Rev.\ Lett.\  {\bf 101}, 142001 (2008).
  [arXiv:0806.0848]
  \href{http://www.slac.stanford.edu/spires/find/hep/www?j=PRLTA,101,142001}
  {[SPIRES]}.

\bibitem{Thaler:2008ju}
  J.~Thaler and L.~T.~Wang,
  {\em Strategies to Identify Boosted Tops},
  JHEP {\bf 0807}, 092 (2008) 
  [arXiv:0806.0023]
  \href{http://www.slac.stanford.edu/spires/find/hep/www?j=JHEPA,0807,092}
  {[SPIRES]}.

\bibitem{Almeida:2008tp}
  L.~G.~Almeida, S.~J.~Lee, G.~Perez, I.~Sung and J.~Virzi,
  {\em Top Jets at the LHC},
  Phys.\ Rev.\  D {\bf 79}, 074012 (2009) 
 [arXiv:0810.0934]
 \href{http://www.slac.stanford.edu/spires/find/hep/www?j=PHRVA,D79,074012}
  {[SPIRES]}.
  
\bibitem{Krohn:2009wm}
  D.~Krohn, J.~Shelton and L.~T.~Wang,
  {\em Measuring the Polarization of Boosted Hadronic Tops},
  arXiv:0909.3855 [hep-ph],
  \href{http://www.slac.stanford.edu/spires/find/hep/www?bb=ARXIV:0909.3855}
  {[SPIRES]}.
  
\bibitem{Plehn:2009rk}
  T.~Plehn, G.~P.~Salam and M.~Spannowsky,
  {\em Fat Jets for a Light Higgs},
  Phys.\ Rev.\ Lett.\  {\bf 104}, 111801 (2010)
  [arXiv:0910.5472]
  \href{http://www.slac.stanford.edu/spires/find/hep/www?j=PRLTA,104,111801}
  {[SPIRES]}.

\bibitem{Mangano:2002ea}
  M.~L.~Mangano, M.~Moretti, F.~Piccinini, R.~Pittau and A.~D.~Polosa,
  {\em ALPGEN, a generator for hard multiparton processes in hadronic
   collisions},
  JHEP {\bf 0307}, 001 (2003).
 [hep-ph/0206293]
 \href{http://www.slac.stanford.edu/spires/find/hep/www?j=JHEPA,0307,001}
  {[SPIRES]}.
  
\bibitem{Sjostrand:2006za}
  T.~Sjostrand, S.~Mrenna and P.~Z.~Skands,
  {\em PYTHIA 6.4 Physics and Manual},
  JHEP {\bf 0605}, 026 (2006) 
  [hep-ph/0603175]
  \href{http://www.slac.stanford.edu/spires/find/hep/www?j=JHEPA,0605,026}
  {[SPIRES]}.
  
\bibitem{Cacciari:2005hq}
  M.~Cacciari and G.~P.~Salam,
  {\em Dispelling the $N^{3}$ myth for the $k_t$ jet-finder},
  Phys.\ Lett.\  B {\bf 641}, 57 (2006) 
  [hep-ph/0512210]
  \href{http://www.slac.stanford.edu/spires/find/hep/www?j=PHLTA,B641,57}
  {[SPIRES]};
  M. Cacciari, G. P. Salam and G. Soyez, 
  \href{http://fastjet.fr}{http://fastjet.fr}.
  
\bibitem{Evans:2009ga}
  J.~A.~Evans and M.~A.~Luty,
  {\em Strong Electroweak Symmetry Breaking and Spin 0 Resonances},
  Phys.\ Rev.\ Lett.\  {\bf 103}, 101801 (2009)
 [arXiv:0904.2182]
 \href{http://www.slac.stanford.edu/spires/find/hep/www?j=PRLTA,103,101801}
  {[SPIRES]}.
  
\bibitem{CAjet}
  Y.~L.~Dokshitzer, G.~D.~Leder, S.~Moretti and B.~R.~Webber,
  {\em Better Jet Clustering Algorithms},
  JHEP {\bf 9708}, 001 (1997)
  \href{http://www.slac.stanford.edu/spires/find/hep/www?j=JHEPA,9708,001}
  {[SPIRES]};
    M.~Wobisch and T.~Wengler,
  {\em Hadronization corrections to jet cross sections in deep-inelastic
   scattering},
  [hep-ph/9907280]
  \href{http://www.slac.stanford.edu/spires/find/hep/www?bb=HEP-PH/9907280}
  {[SPIRES]}.

\bibitem{antiKT}
  M.~Cacciari, G.~P.~Salam and G.~Soyez,
  {\em The anti-$k_t$ jet clustering algorithm},
  JHEP {\bf 0804}, 063 (2008) 
 [arXiv:0802.1189]
\href{http://www.slac.stanford.edu/spires/find/hep/www?j=JHEPA,0804,063}
  {[SPIRES]}.
  
\bibitem{KTjet}
  S.~D.~Ellis and D.~E.~Soper,
  {\em Successive combination jet algorithm for hadron collisions},
  Phys.\ Rev.\  D {\bf 48}, 3160 (1993).
  [arXiv:hep-ph/9305266]
  \href{http://www.slac.stanford.edu/spires/find/hep/www?j=PHRVA,D48,3160}
  {[SPIRES]}.
  
\bibitem{Kribs:2009yh}
  G.~D.~Kribs, A.~Martin, T.~S.~Roy and M.~Spannowsky,
  {\em Discovering the Higgs Boson in New Physics Events using Jet Substructure},
  [arXiv:0912.4731]
  \href{http://www.slac.stanford.edu/spires/find/hep/www?bb=ARXIV:0912.4731}
  {[SPIRES]}.
  
\bibitem{btags}
  ATLAS Collaboration, 
  {\em ATLAS Sensitivity to the Standard Model Higgs in the HW and HZ 
  Channels at High Transverse Momenta},  
  \href{http://cdsweb.cern.ch/record/1201444/}
  {ATL-PHYS-PUB-2009-088}.
  
\bibitem{smearing}
  A. Heister et al. [CMS Collaboration], CMS Note 2006/036 
  \href{http://www.slac.stanford.edu/spires/find/hep/www?r=CERN-CMS-NOTE-2006-036}
  {[SPIRES]}.

\bibitem{GU:Stat}
  J.~F.~Gunion and D.~E.~Soper,
  {\em Statistical analysis in new particle searches},
  Phys.\ Rev.\  D {\bf 35} (1987) 179
  \href{http://www.slac.stanford.edu/spires/find/hep/www?j=PHRVA,D35,179}
  {[SPIRES]}.

\end{thebibliography}
\end{document}